\documentclass[aps, prd, amsmath, floats, floatfix, twocolumn, nofootinbib, superscriptaddress]{revtex4-1}
\usepackage[pdftex]{graphicx}
\usepackage{color}
\usepackage{latexsym}
\usepackage[colorlinks]{hyperref}
\usepackage{amsthm,amsmath,amssymb}
\usepackage{mathrsfs}

\newcommand{\be}{\begin{eqnarray}}
\newcommand{\ee}{\end{eqnarray}}

\begin{document}

\title{Testing Black Holes with Interstellar Missions: II. Flyby Probes}

\author{Yi~Fan}
\affiliation{Center for Astronomy and Astrophysics, Department of Physics, Fudan University, Shanghai 200438, China}

\author{Cosimo~Bambi}
\email[Corresponding author: ]{bambi@fudan.edu.cn}
\affiliation{Center for Astronomy and Astrophysics, Department of Physics, Fudan University, Shanghai 200438, China}
\affiliation{School of Natural Sciences and Humanities, New Uzbekistan University, Tashkent 100000, Uzbekistan}

\author{Leda~Gao}
\affiliation{Center for Astronomy and Astrophysics, Department of Physics, Fudan University, Shanghai 200438, China}

\author{Abdurakhmon~Nosirov}
\affiliation{Center for Astronomy and Astrophysics, Department of Physics, Fudan University, Shanghai 200438, China}
\affiliation{Institut f\"ur Astronomie und Astrophysik, Eberhard-Karls Universit\"at T\"ubingen, D-72076 T\"ubingen, Germany}

\author{Andrea~Santangelo}
\affiliation{Institut f\"ur Astronomie und Astrophysik, Eberhard-Karls Universit\"at T\"ubingen, D-72076 T\"ubingen, Germany}
\affiliation{Center for Astronomy and Astrophysics, Department of Physics, Fudan University, Shanghai 200438, China}

\begin{abstract}
Recently, we demonstrated that while an interstellar mission to the nearest black hole remains highly speculative and extraordinarily challenging, it is not entirely implausible within the coming decades. Given that such a mission would likely take about a hundred years and require substantial financial and human investment, it is essential to assess whether it could investigate black holes and test General Relativity to a degree that cannot be achieved by Solar System observatories for the foreseeable future. In Paper~I, we assumed the capability to decelerate the spacecraft and presented a preliminary study of how orbiting probes could test the nature of the compact object. In this second paper, we study how the black hole can be tested without decelerating the spacecraft, using flyby probes.
\end{abstract}

\maketitle

%%%%%%%%%%%%%%%%% BODY OF PAPER %%%%%%%%%%%%%%%%%%

\section{Introduction}

The idea of developing spacecraft for missions beyond the Solar System dates back to the Orion project in the 1950s~\cite{Orion} and the Daedalus project in the 1970s~\cite{Daedalus}. In the 1980s, attention shifted from nuclear propulsion toward beamed propulsion, which was first suggested in the 1960s~\cite{Marx,Redding} and is now regarded as the most viable option for the relatively short term. Over the last 10-20 years, the exoplanet community has frequently considered interstellar voyages to examine exoplanets in neighboring star systems~\cite{Lubin16,Lubin22,Parkin18,Kuhlmey25,Eubanks2026}. The most well-known project was the Breakthrough Starshot Initiative,\footnote{\href{https://breakthroughinitiatives.org/}{https://breakthroughinitiatives.org/}} which sought to develop a tiny spacecraft capable of traveling at one-fifth the speed of light, reaching Alpha Centauri in roughly 20 years, but was discontinued in September 2025.

In current interstellar mission concepts, the spacecraft -- often referred to as a nanocraft because of its extremely low mass -- consists of a gram-sized wafer (which serves as a fully operational probe containing a computer processor, navigation, and communication systems) along with an exceptionally thin, meter-scale dielectric metamaterial light sail. High-power lasers located on the ground or in space strike the light sail, and the resulting radiation pressure accelerates the nanocraft to its intended speed. No fundamental technical barriers prevent reaching 90\% of the speed of light using this approach, although higher velocities immediately increase mission costs.

In Refs.~\cite{Bambi:2025kcr,Bambi:2025hjn,Bambi:2026dwc}, we examined the prospect of sending some nanocrafts to the nearest black hole, aiming to test black holes and General Relativity at levels likely unattainable through astrophysical observations from Earth or within the Solar System; see, for instance, Refs.~\cite{Bambi:2017khi,Bambi:2015kza,Yagi:2016jml}. Basic estimates suggest that a few black holes could reside within 50 light-years of the Solar System, even though detecting them is quite difficult~\cite{Murchikova:2025oio,Nosirov:2026fjo,Nosirov2}. If a black hole is found 20-25 light-years away and a fleet of nanocrafts is launched at one-third the speed of light, the nanocrafts could reach the black hole in 60-75 years, carry out various scientific investigations to probe the nature of the compact object and the behavior of strong gravitational fields, and transmit all data back to Earth. The data would take another 20-25 years to return, making the overall mission duration roughly 80-100 years. If the black hole is closer or farther, the mission duration scales down or up accordingly. If the nanocrafts can achieve a higher speed (requiring either more sophisticated technology or a larger mission budget), the mission duration decreases.

Since such a mission to the nearest black hole might last about a century and require considerable economic and personnel resources, it is crucial to assess whether it could truly explore black holes and test General Relativity at a level that observatories inside the Solar System will be unable to reach for many years. In Ref.~\cite{Gao:2026jpl} (Paper~I), we began the study of what kind of measurements should be performed near the black hole to test the nature of the compact object and the physics of strong gravitational fields. We assumed that these nanocrafts can decelerate and enter certain orbits, and therefore we investigated how {\it orbiting probes} can test the spacetime metric around the black hole and the existence of the event horizon. The conclusion of Paper~I is that the ability to decelerate the probes is not sufficient. To perform very stringent tests of the spacetime geometry around a black hole, we need at least one probe to orbit very close to the innermost stable circular orbit.

In the present manuscript (Paper~II), we study how to test the nature of the black hole without decelerating the nanocrafts; that is, with {\it flyby probes}. As in Paper~I, we assume a simplified situation where the black hole does not rotate and is surrounded by vacuum. In reality, the black hole would possess some spin angular momentum and would be drawing matter from the interstellar medium. Although the medium's density is low, it increases as we get closer to the black hole, and its effect on the motion and observations of our probes must be carefully evaluated. Moreover, as in Paper~I, the goal here is to study which kind of measurements should be performed by the nanocrafts in order to test General Relativity with a certain precision. We do not assess the feasibility of developing the necessary technology -- a task that is certainly very important for designing a real mission and is left to future studies.

The content of the manuscript is as follows. In Sections~\ref{s-uco} and \ref{s-zwm}, we consider the Schwarzschild spacetime: first, we review the calculations of the critical impact parameter separating scattered orbits from captured orbits, and then we study the motion of massive particles with impact parameters close to the critical one. In Section~\ref{s-1}, we study the motion of massive particles with impact parameters close to the critical one in a deformed Schwarzschild spacetime. In Section~\ref{s-2}, we compare the motion in the Schwarzschild and deformed Schwarzschild spacetimes. In Section~\ref{s-c}, we discuss how we can measure the mass and possible deviations from the Schwarzschild metric using a swarm of probes.

%%%%%%%%%%%%%%%%%%%%%%%%%%%%%%%%%%%%%%%%%%%%%%%%%% 

\section{Critical Impact Parameter}\label{s-uco}

A flyby experiment is a space mission where a spacecraft passes close to the target source (in our case, a black hole) without entering its orbit. In this section, we assume a Schwarzschild background and calculate the critical impact parameter that separates trajectories scattered back to infinity from those captured by the black hole, as a function of the asymptotic velocity of the probe.

In the Schwarzschild coordinates $(t,r,\theta,\phi)$, the line element reads
\begin{eqnarray}
ds^2 &=&
-\left(1-\frac{2M}{r}\right)dt^2
+\left(1-\frac{2M}{r}\right)^{-1}dr^2
\nonumber\\
&&
+r^2 d\theta^2
+r^2\sin^2\theta d\phi^2 ,
\end{eqnarray}
where $M$ is the black hole mass parameter. Here and throughout the manuscript, we adopt natural units in which $G_{\rm N}=c=1$ and the metric signature $(-+++)$. However, for the numerical simulations of nanocraft trajectories near black holes, it proves more practical to express the black hole mass in solar mass units $M_\odot$, and to measure spatial separations between trajectories and time differences between nanocraft-borne clocks in centimeters and seconds, respectively. This choice is motivated by the desire to facilitate future assessments of the technological requirements for such an interstellar mission.

We choose a coordinate system such that the trajectory of the probe lies on the equatorial plane. Stationarity and axisymmetry give two constants of motion: the specific energy $E$ and the specific angular momentum $L$ of the probe,
\begin{equation}
E=\left(1-\frac{2M}{r}\right)\dot t,\qquad
L=r^2\dot\phi ,
\end{equation}
where a dot denotes a derivative with respect to the probe's proper time. Using the normalization of the four-velocity, $g_{\mu\nu}\dot x^\mu\dot x^\nu=-1$, the radial equation can be written as
\begin{equation}
\dot r^2=E^2-V_{\rm eff}(r;L),
\label{eq:radial-sch}
\end{equation}
with
\begin{equation}
V_{\rm eff}(r;L)=
\left(1-\frac{2M}{r}\right)
\left(1+\frac{L^2}{r^2}\right).
\label{eq:veff-sch}
\end{equation}
In contrast with the Newtonian problem, the effective potential includes the attractive term $-2ML^2/r^3$, which lowers the centrifugal barrier at small radii and allows the capture of probes with non-vanishing angular momentum. The extrema of $V_{\rm eff}$ are determined by $M r^2-L^2 r+3ML^2=0$, whose discriminant is $L^2(L^2-12M^2)$. For $L>2\sqrt{3}\,M$, $V_{\rm eff}$ has a local maximum and a local minimum; the maximum corresponds to an unstable circular orbit.

For a probe arriving from infinity with velocity $v_\infty$, the specific energy is
\begin{equation}
E=\gamma_\infty=\frac{1}{\sqrt{1-v_\infty^2}} ,
\end{equation}
where $\gamma_\infty$ is the probe's Lorentz factor at infinity. If $b$ is the impact parameter of the trajectory, namely the perpendicular distance between the black hole and the straight line that the probe would follow in the absence of gravitational deflection, the specific angular momentum is $L=\gamma_\infty v_\infty b$, where $\gamma_\infty v_\infty$ is the probe's specific linear momentum at infinity. The impact parameter is thus
\begin{equation}
b=\frac{L}{\gamma_\infty v_\infty}
=\frac{L}{\sqrt{E^2-1}} .
\label{eq:impact-parameter}
\end{equation}

For fixed $v_\infty$, the height of the centrifugal barrier decreases as the impact parameter decreases. The trajectory exactly at the capture/escape threshold has an energy matching the top of the barrier and asymptotically approaches the unstable circular orbit at the barrier maximum. If $r_{\rm c}$ is the radius of this orbit, we have
\begin{equation}
E^2_{\rm c}=V_{\rm eff}(r_{\rm c};L_{\rm c}),
\qquad
V_{\rm eff}'(r_{\rm c};L_{\rm c})=0 ,
\label{eq:critical-conditions}
\end{equation}
where the prime denotes a derivative with respect to $r$. These conditions give
\begin{equation}
L_{\rm c}^2=\frac{M r_{\rm c}^2}{r_{\rm c}-3M},
\qquad
E_{\rm c}^2=
\frac{(r_{\rm c}-2M)^2}{r_{\rm c}(r_{\rm c}-3M)} .
\label{eq:circular-energy-angular}
\end{equation}
Moreover,
\begin{equation}
V_{\rm eff}''(r_{\rm c};L_{\rm c})
=
\frac{2M(r_{\rm c}-6M)}
{r_{\rm c}^3(r_{\rm c}-3M)} ,
\label{eq:vpp-sch}
\end{equation}
so circular orbits are unstable for $3M \le r_{\rm c}<6M$ and stable for $r_{\rm c} \ge 6M$. Since trajectories coming from infinity have $E\geq 1$ ($E=1$ for $v_\infty=0$), the relevant unstable circular orbits have radii $3M \le r_{\rm c} \le 4M$.

Imposing $E_{\rm c}=\gamma_\infty$ in Eq.~(\ref{eq:circular-energy-angular}), we find that $x_{\rm c}=r_{\rm c}/M$ satisfies the quadratic equation
\begin{equation}
v_\infty^2 x_{\rm c}^2
+\left(1-4v_\infty^2\right)x_{\rm c}
-4\left(1-v_\infty^2\right)=0 ,
\end{equation}
whose physical root ($x_{\rm c}>3$) gives the critical radius as a function of the velocity at infinity
\begin{equation}
\frac{r_{\rm c}}{M}=
\frac{4v_\infty^2-1+\sqrt{1+8v_\infty^2}}
{2v_\infty^2}.
\label{eq:rc-vinf}
\end{equation}
We recover the expected limits: $r_{\rm c}\rightarrow 4M$ for $v_\infty\rightarrow 0$ and $r_{\rm c}\rightarrow 3M$ for $v_\infty\rightarrow 1$. From Eqs.~(\ref{eq:impact-parameter}) and (\ref{eq:circular-energy-angular}), the critical impact parameter is
\begin{equation}
b_{\rm crit}^2(v_\infty)=
\frac{L_{\rm c}^2}{E_{\rm c}^2-1}=
\frac{r_{\rm c}^3}{4M-r_{\rm c}} .
\label{eq:bcrit-vinf}
\end{equation}
For $v_\infty\rightarrow 1$, $b_{\rm crit}\rightarrow 3\sqrt{3}\,M$ and we recover the critical impact parameter of photons; for
$v_\infty\rightarrow 0$, $b_{\rm crit}\approx 4M/v_\infty$, which diverges. For fixed $v_\infty$, probes with $b>b_{\rm crit}$ are scattered back to infinity, while those with $b<b_{\rm crit}$ plunge into the black hole. In Tab.~\ref{t-critical}, we report the values of $r_{\rm c}$ and $b_{\rm crit}$ for some representative values of $v_\infty$. Trajectories with impact parameters close to $b_{\rm crit}$ are the subject of the next section.

\begin{table}[t]
\caption{Critical radius $r_{\rm c}$ and critical impact parameter $b_{\rm crit}$ for different values of the probe asymptotic velocity $v_\infty$.}
\label{t-critical}
\begin{ruledtabular}
\begin{tabular}{ccc}
$v_\infty$ & $r_{\rm c}/M$ & $b_{\rm crit}/M$\\
\hline
0.1 & 3.962 & 40.198\\
0.2 & 3.861 & 20.382\\
0.3 & 3.730 & 13.879\\
0.5 & 3.464 & 8.807\\
0.8 & 3.151 & 6.073
\end{tabular}
\end{ruledtabular}
\end{table}

%%%%%%%%%%%%%%%%%%%%%%%%%%%%%%%%%%%%%%%%%%%%%%%%%%

\section{Zoom-Whirl Motion}\label{s-zwm}

In this section, we study the motion of probes with impact parameters close to the critical value $b_{\rm crit} = b_{\rm crit}(v_\infty)$. If the initial conditions are sufficiently close to the capture/escape threshold, the probe approaches the unstable circular orbit at $r=r_{\rm c}$, revolves around the black hole near this orbit for some time, and eventually either escapes back to infinity or plunges into the black hole. During the near-circular phase, the probe can accumulate an azimuthal angle largely exceeding $2\pi$. This is the so-called {\it zoom-whirl motion}~\cite{Glampedakis:2002ya}.

The logarithmic growth of the azimuthal angle near the capture/escape threshold can be derived analytically. Let $\lambda$ be the parameter that is varied across the threshold. We may take $\lambda=b/M$ at fixed $v_\infty$ or $\lambda=v_\infty$ at fixed $b$, so $\lambda$ is dimensionless. We write $\lambda = \lambda_{\rm c} + \Delta$, where $\lambda_{\rm c}$ is the critical parameter, so that $\Delta>0$ on the scattered side and $\Delta<0$ on the captured side.

It is useful to write the radial equation as $\dot r^2=R(r;\lambda)$, where $R(r;\lambda)=E^2(\lambda)-V_{\rm eff}[r;L(\lambda)]$. For $\lambda = \lambda_{\rm c}$, we have $R(r_{\rm c};\lambda_{\rm c})=0$ and $\partial_rR(r_{\rm c};\lambda_{\rm c})=0$, see Eq.~(\ref{eq:critical-conditions}). We define the positive constant $\kappa$ by $R(r_{\rm c};\lambda)=-\kappa\Delta+O(\Delta^2)$, so that the expansion around $(r_{\rm c},\lambda_{\rm c})$ gives
\begin{equation}
\dot r^2\simeq
-\kappa\Delta
- \frac{1}{2} V_{\rm eff}''(r_{\rm c};L_{\rm c}) \, \delta r^2 ,
\label{eq:near-critical-radial}
\end{equation}
where $\delta r = r - r_{\rm c}$ and $V_{\rm eff}''(r_{\rm c};L_{\rm c})<0$ for unstable circular orbits. On the scattered side ($\Delta>0$), the probe reaches the turning point
\begin{equation}
\delta r_{\rm tp}=\sqrt{\frac{2\kappa\Delta}{-V_{\rm eff}''(r_{\rm c};L_{\rm c})}} ,
\end{equation}
just outside the unstable circular orbit. On the captured side ($\Delta<0$), there is no turning point and $\dot r^2\simeq -\kappa \Delta > 0$ at $r_{\rm c}$. In both cases, the radial motion near $r_{\rm c}$ is slow, while the azimuthal motion proceeds at the finite rate $\dot\phi\simeq L_{\rm c}/r_{\rm c}^2$.

For escaping trajectories, the experimentally cleaner angular variable is the total azimuthal change between two large-radius crossings,
\begin{equation}
\Delta\phi_{\rm tot}
=|\phi_{\rm out}(R_{\rm m})-\phi_{\rm in}(R_{\rm m})| ,
\label{eq:dphitot-def}
\end{equation}
where $\phi_{\rm in}$ is evaluated on the ingoing branch and $\phi_{\rm out}$ on the outgoing branch at the same large matching radius $R_{\rm m}$. The angular integral can be separated into a regular part and a singular part. The regular part remains finite as $\Delta\rightarrow0$ and depends on the finite-radius matching convention only through an additive matching constant. The singular part is generated by the slow radial motion near $r_{\rm c}$. Using Eq.~(\ref{eq:near-critical-radial}), this contribution is
\begin{equation}
\Delta\phi_{\rm sing} = \int \frac{\dot{\phi}}{\dot{r}} dr 
\simeq
\frac{L_{\rm c}}{r_{\rm c}^2}
\int\frac{d(\delta r)}
{\sqrt{-\kappa\Delta
- V_{\rm eff}''(r_{\rm c};L_{\rm c}) \delta r^2/2}} .
\label{eq:dphi-integral}
\end{equation}
Using Eqs.~(\ref{eq:circular-energy-angular}) and (\ref{eq:vpp-sch}), we have
\begin{equation}
\frac{r_{\rm c}^4}{2L_{\rm c}^2}
V_{\rm eff}''(r_{\rm c};L_{\rm c})
=
\frac{r_{\rm c} - 6M}{r_{\rm c}} ,
\end{equation}
and the integral in Eq.~(\ref{eq:dphi-integral}) is elementary. For escaping trajectories in the limit $\Delta\rightarrow 0^+$, the total azimuthal change obeys
\begin{equation}\label{eq:dphitot-log}
\Delta\phi_{\rm tot}
\simeq
\sqrt{\frac{r_{\rm c}}{6M-r_{\rm c}}}
\left(-\ln |\Delta|+C_\phi\right) ,
\end{equation}
where $C_\phi$ is a finite matching constant. It absorbs $\kappa$, the normalization of $\Delta$, and the regular contribution outside the near-separatrix region. For a finite matching radius, this constant also contains the corresponding finite-radius correction; in the asymptotic definition it approaches an $R_{\rm m}$-independent value once $R_{\rm m}$ is sufficiently large. Thus the logarithmic slope is fixed by the local expansion around the unstable orbit, while the actual boundary location and the finite constant are obtained from numerical trajectories. The factor $\sqrt{r_{\rm c}/\left(6M-r_{\rm c}\right)}$ depends only on the radius of the unstable circular orbit: it decreases from $\sqrt{2}$ for $r_{\rm c}=4M$ ($v_\infty\rightarrow 0$) to 1 for $r_{\rm c}=3M$ ($v_\infty\rightarrow 1$).

\begin{figure}[tbp]
\centering
\includegraphics[width=0.9\linewidth]{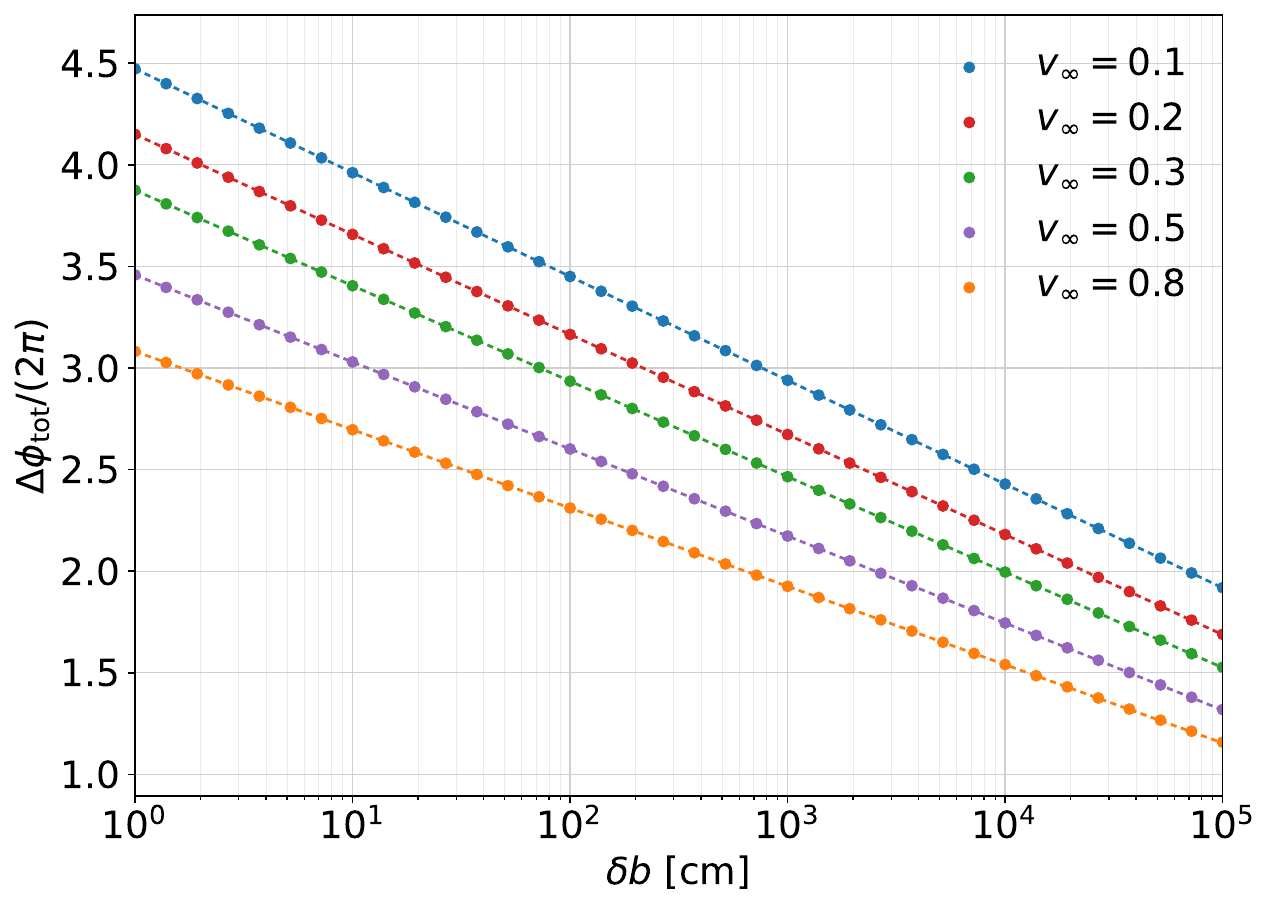}
%\vspace{0.0cm}
\caption{Observable total azimuthal change $\Delta\phi_{\rm tot}/(2\pi)$ as a function of the deviation from the critical impact parameter $\delta b = b - b_{\rm crit}$ on the escaping side, $\delta b>0$, for different values of the probe asymptotic velocity $v_\infty$: $v_\infty = 0.1$ ($b_{\rm crit} = 40.198$~$M$), $v_\infty = 0.2$ ($b_{\rm crit} \approx 20.382$~$M$), $v_\infty = 0.3$ ($b_{\rm crit} \approx 13.879$~$M$), $v_\infty = 0.5$ ($b_{\rm crit} \approx 8.807$~$M$), and $v_\infty = 0.8$ ($b_{\rm crit} \approx 6.073$~$M$). The angle is evaluated between the ingoing and outgoing crossings of the finite matching radius $R_{\rm m}=50M$. We assume a black hole mass $M = 10$~$M_\odot$.
\label{f-nw1}}
\end{figure}

\begin{figure}[tbp]
\centering
\includegraphics[width=0.9\linewidth]{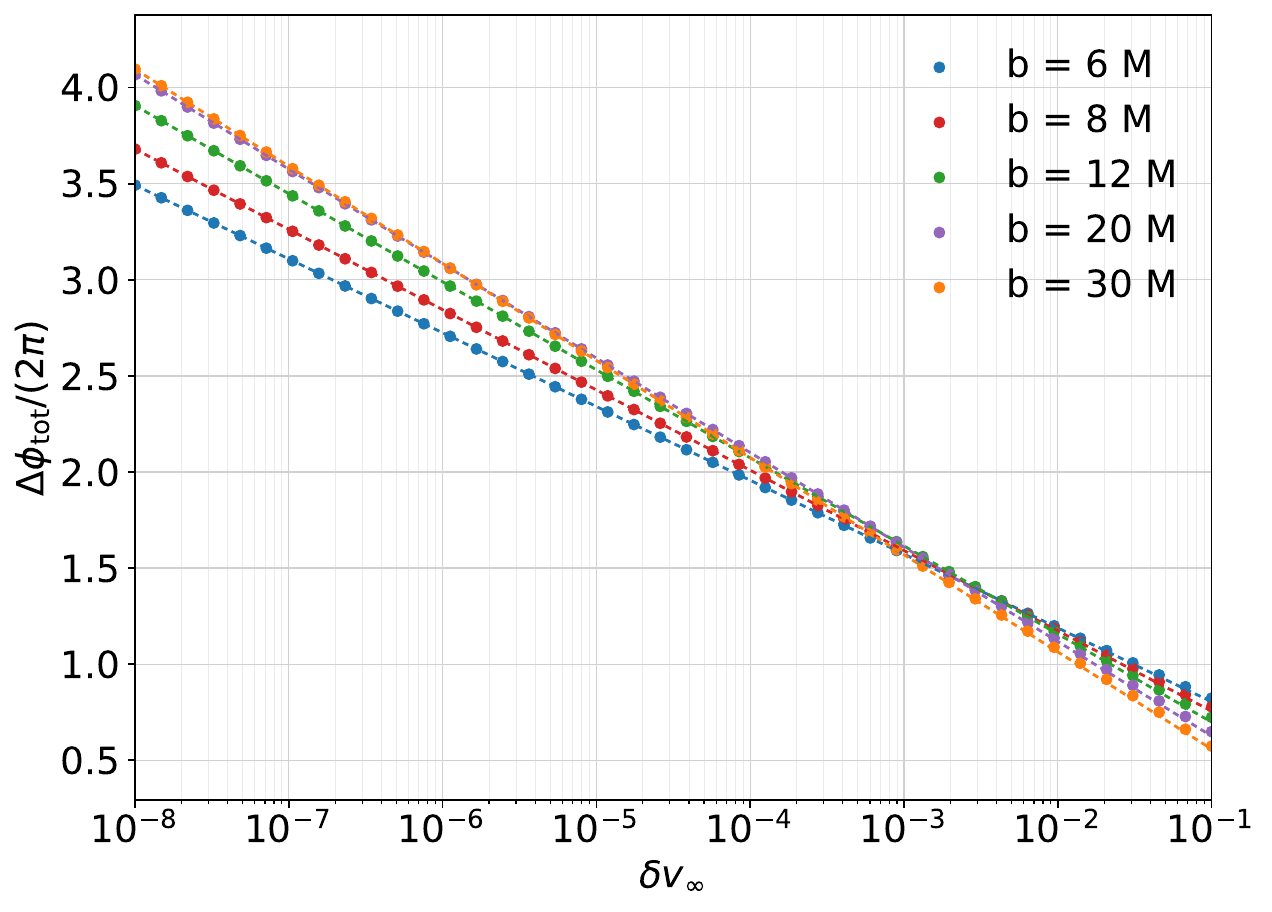}
%\vspace{0.0cm}
\caption{Observable total azimuthal change $\Delta\phi_{\rm tot}/(2\pi)$ as a function of the deviation from the critical asymptotic velocity $\delta v_\infty = v_\infty - v_{\rm crit}$ on the escaping side, $\delta v_\infty>0$, for different values of the impact parameter $b$: $b = 6$~$M$ ($v_{\rm crit} \approx 0.813$), $b = 8$~$M$ ($v_{\rm crit} \approx 0.561$), $b = 12$~$M$ ($v_{\rm crit} \approx 0.352$), $b = 20$~$M$ ($v_{\rm crit} \approx 0.204$), and $b = 30$~$M$ ($v_{\rm crit} \approx 0.135$). The angle is evaluated between the ingoing and outgoing crossings of the finite matching radius $R_{\rm m}=50M$.
\label{f-nw2}}
\end{figure}

\begin{figure}[tbp]
\centering
\includegraphics[width=0.9\linewidth]{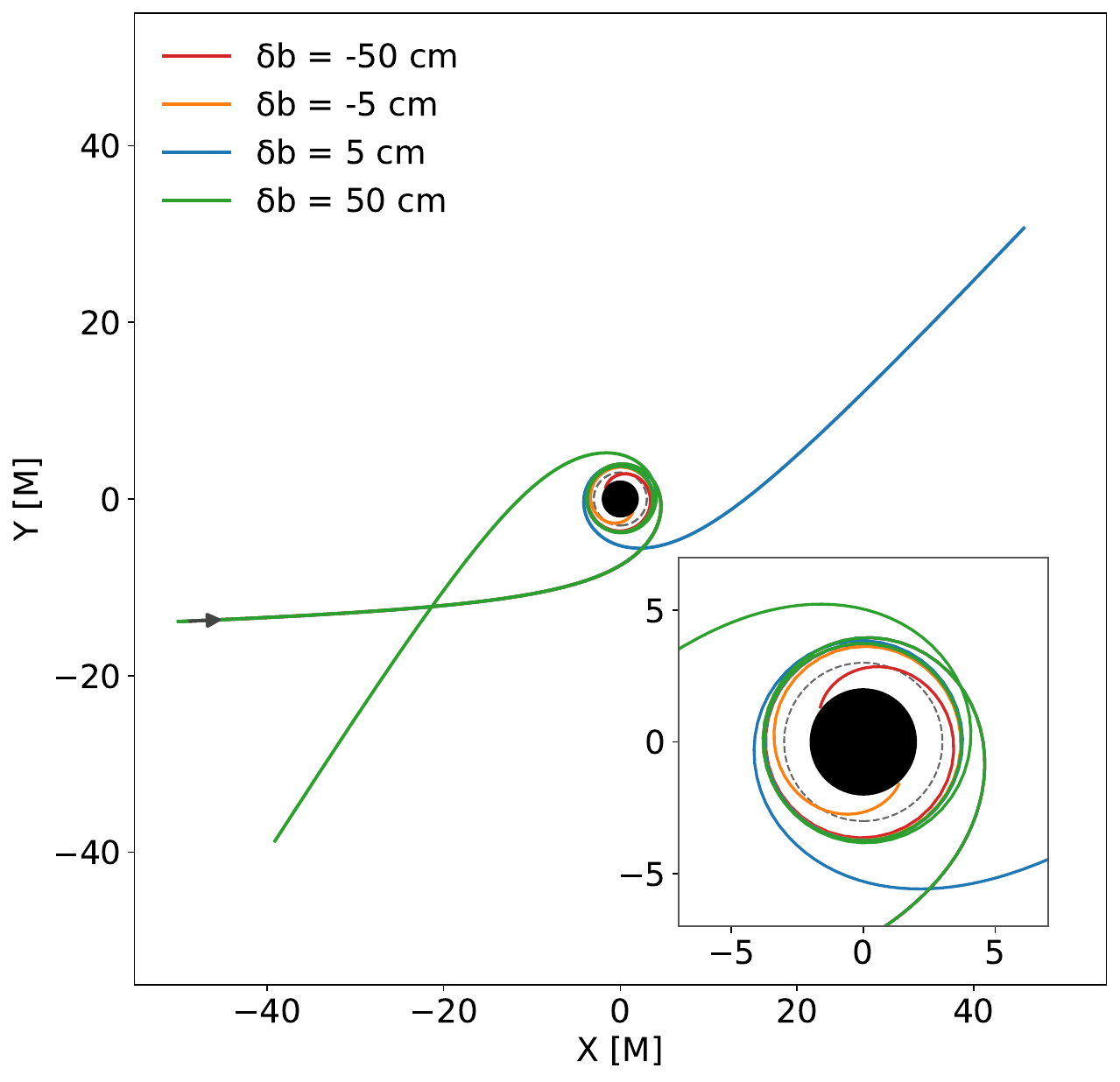}
%\vspace{0.0cm}
\caption{Trajectories around a Schwarzschild black hole with $M = 10$~$M_\odot$ for $\delta b = -50$~cm, $-5$~cm, 5~cm, and 50~cm. The asymptotic velocity is $v_\infty = 0.3$. The black arrow indicates the propagation direction of the probes. The black-dashed circle in the small quadrant marks the photon radius. 
\label{f-sch}}
\end{figure}

To verify Eq.~(\ref{eq:dphitot-log}), we integrate numerically the geodesic equations for probes launched from the large matching radius $R_{\rm m}=50M$ with asymptotic velocity $v_\infty$ and impact parameter $b$. For escaping trajectories, the integration stops at the outgoing crossing of the same radius and we compute $\Delta\phi_{\rm tot}$ from Eq.~(\ref{eq:dphitot-def}). 
In Fig.~\ref{f-nw1}, we show $\Delta\phi_{\rm tot}/(2\pi)$ as a function of $\delta b=b-b_{\rm crit}(v_\infty)$ on the escaping side for a black hole with $M=10~M_\odot$ and different values of $v_\infty$. The numerical results (dots) follow the logarithmic law in Eq.~(\ref{eq:dphitot-log}) (dashed lines). The predicted slope per decade of $\delta b$ in $\Delta\phi_{\rm tot}/(2\pi)$ is $\left(\ln 10/2\pi\right)\sqrt{r_{\rm c}/\left(6M-r_{\rm c}\right)}$, which ranges from 0.51 for $v_\infty=0.1$ to 0.39 for $v_\infty=0.8$, in agreement with the numerical results. Captured probes do not have an outgoing large-radius endpoint; they contribute only to bracketing the capture boundary.

We can also approach the capture/escape threshold by varying the asymptotic velocity at fixed impact parameter. Since
$b_{\rm crit}(v_\infty)$ decreases monotonically with $v_\infty$ (see Tab.~\ref{t-critical}), for every $b>3\sqrt{3}\,M$ there is a critical velocity $v_{\rm crit}(b)$, obtained by inverting Eq.~(\ref{eq:bcrit-vinf}): probes with $v_\infty>v_{\rm crit}$ are
scattered back to infinity, while probes with $v_\infty<v_{\rm crit}$ are captured by the black hole. In Fig.~\ref{f-nw2}, we show $\Delta\phi_{\rm tot}/(2\pi)$ as a function of $\delta v_\infty=v_\infty-v_{\rm crit}(b)$ on the escaping side for different values of $b$. As in Fig.~\ref{f-nw1}, the numerical results follow the logarithmic law in Eq.~(\ref{eq:dphitot-log}).

In Fig.~\ref{f-sch}, we show four representative trajectories around a black hole with $M=10~M_\odot$ for $v_\infty=0.3$, corresponding to $b_{\rm crit}\approx 13.879\,M$. The four trajectories have $\delta b=\pm 5$~cm and $\pm 50$~cm. Since $M\simeq 14.77$~km for a $10~M_\odot$ black hole, these values correspond to $|\delta b|/M\sim 3\cdot 10^{-6}$ and $\sim 3\cdot 10^{-5}$, respectively. The four probes follow trajectories that are indistinguishable until they reach the near-separatrix region, where they revolve around the black hole slightly outside the photon radius ($r_{\rm c}\approx 3.73\,M$, while the black-dashed circle in the small quadrant of Fig.~\ref{f-sch} marks $r=3M$). After this phase, the two probes with $\delta b>0$ escape back to infinity along two very different directions, while the two probes with $\delta b<0$ plunge into the black hole. As expected from Eq.~(\ref{eq:dphitot-log}), smaller values of $|\delta b|$ produce larger accumulated azimuthal angles. Since the growth is only logarithmic as $\delta b\rightarrow 0$, every additional revolution around the black hole requires reducing $|\delta b|$ by a factor $e^{2\pi\sqrt{\left(6M-r_{\rm c}\right)/r_{\rm c}}}\approx 134$ (for $v_\infty=0.3$): ten whirl-like revolutions would already require fine-tuning the impact parameter below the size of an atomic nucleus.

%%%%%%%%%%%%%%%%%%%%%%%%%%%%%%%%%%%%%%%%%%%%%%%%%%

\section{Deformed Black Holes}\label{s-1}

Sections~\ref{s-uco} and \ref{s-zwm} assumed a Schwarzschild geometry and showed that flyby probes with impact parameters close to the critical value can display zoom-whirl motion before either escaping to infinity or plunging into the black hole. We now replace the Schwarzschild metric by a deformed Schwarzschild metric and compute how the capture boundary and near-critical motion change. We consider the Johannsen spacetime with the deformation parameter $\alpha_{13}$~\cite{Johannsen:2013szh}. For $\alpha_{13}=0$, we recover the Schwarzschild solution; a non-vanishing value of $\alpha_{13}$ parametrizes a deviation from the Schwarzschild geometry. Current observational constraints on $\alpha_{13}$ from X-ray and gravitational-wave data are summarized in Paper~I~\cite{Gao:2026jpl} and references therein~\cite{Tripathi:2020yts,Das:2026zyt}; their present scale is of order $|\alpha_{13}|\lesssim 0.1$, while future facilities such as the space interferometer LISA are expected to improve this to $|\alpha_{13}|\lesssim 0.01$. The benchmark value $\alpha_{13}=10^{-5}$ adopted in this section lies orders of magnitude below these scales; for a stellar-mass black hole it is nonetheless large enough to displace the capture/escape boundary by centimeters. Resolving such a shift would require centimeter-scale control of the probes' asymptotic data.

Specializing the Johannsen spacetime to the case of vanishing spin angular momentum and retaining only the deformation parameter $\alpha_{13}$, the line element reads~\cite{Johannsen:2013szh}
\be\label{eq-j}
ds^2 &=& - \left( 1 - \frac{2 M}{r} \right) 
\left( 1 + \alpha_{13} \frac{M^3}{r^3} \right)^{-2} dt^2 \nonumber\\
&& + \left( 1 - \frac{2 M}{r} \right)^{-1} dr^2 
+ r^2 d\theta^2 + r^2 \sin^2\theta d\phi^2 \, . \qquad
\ee
We acknowledge that the Johannsen spacetime parametrized by the deformation parameter $\alpha_{13}$ does not exhaust all physically viable static and spherically symmetric departures from the Schwarzschild solution. Consequently, our scrutiny of the Schwarzschild metric is inherently model-dependent. Nonetheless, as demonstrated in Ref.~\cite{Zhao:2023erf}, such tests remain a valuable probe for new physics, even if their discriminatory power may be less than that achievable with an exact black hole metric. Moreover, the primary objective of the present work is to explore the potential of nanocraft-based measurements for probing black hole nature, and to identify which observables are best suited for this purpose. Our findings are indeed intended to serve as a foundation for subsequent evaluations of the technological feasibility of such an ambitious interstellar mission.

Let us introduce the functions $f$ and $h$
\begin{equation}
f=1-\frac{2M}{r},\qquad
h=1+\alpha_{13}\frac{M^3}{r^3}.
\end{equation}
For equatorial timelike geodesics in the spacetime of Eq.~(\ref{eq-j}), the conserved specific energy and angular momentum
are
\begin{equation}
E=\frac{f}{h^2}\dot t,\qquad L=r^2\dot\phi ,
\end{equation}
and the normalization of the four-velocity gives
\begin{equation}
\dot r^2=
E^2h^2
-f\left(1+\frac{L^2}{r^2}\right).
\label{eq:radial-joh}
\end{equation}
At a radial turning point, Eq.~(\ref{eq:radial-joh}) can be written as
\begin{equation}
E^2=
\frac{
\left(1-\frac{2M}{r}\right)
\left(1+\frac{L^2}{r^2}\right)}
{\left(1+\alpha_{13}M^3/r^3\right)^2}.
\label{eq:turning-joh}
\end{equation}
We compare the Schwarzschild and Johannsen metrics at fixed asymptotic data. Since $h\rightarrow 1$ at infinity, the asymptotic quantities used in the Schwarzschild calculation have the same operational meaning in the two spacetimes:
\begin{equation}
E=\gamma_\infty=\frac{1}{\sqrt{1-v_\infty^2}},
\qquad
L=\gamma_\infty v_\infty b .
\end{equation}
We therefore compare trajectories with the same physical impact parameter $b$ and the same asymptotic velocity $v_\infty$. These conditions fix $E$ and $L$ in both metrics, while the radial potential, the near-black hole trajectory, and the position of the capture/escape boundary depend on the spacetime geometry.

For fixed $v_\infty$, the critical impact parameter is obtained by requiring that the turning point in Eq.~(\ref{eq:turning-joh}) be at the maximum of the effective barrier. If the right-hand side of Eq.~(\ref{eq:turning-joh}) is denoted by $V_{\rm eff}(r,L,\alpha_{13})$, the critical trajectory satisfies
\begin{equation}
E^2=V_{\rm eff}(r_{\rm c},L,\alpha_{13}),
\qquad
\partial_r V_{\rm eff}(r_{\rm c},L,\alpha_{13})=0 .
\label{eq:critical-joh}
\end{equation}
Expanding the solution to first order in $\alpha_{13}$ (Appendix~\ref{app:joh}) gives
\begin{equation}
b_{\rm crit}(v_\infty,\alpha_{13})
=M\left[\beta_0(v_\infty)+\beta_1(v_\infty)\alpha_{13}
+O(\alpha_{13}^2)\right],
\label{eq:bcrit-joh}
\end{equation}
where
\begin{equation}
\beta_0=
\frac{x_{\rm c}^{\rm Schw}\sqrt{1-v_\infty^2}}
{v_\infty\sqrt{x_{\rm c}^{\rm Schw}-3}},
\qquad
\beta_1=\frac{1}{v_\infty (x_{\rm c}^{\rm Schw})^{3/2}},
\label{eq:kappa-joh}
\end{equation}
and $x_{\rm c}^{\rm Schw}=r_{\rm c}^{\rm Schw}/M$ is the Schwarzschild critical radius in units of $M$, given by Eq.~(\ref{eq:rc-vinf}). Using Eq.~(\ref{eq:rc-vinf}), one may rewrite $v_\infty^2=(4-x_{\rm c}^{\rm Schw})/(x_{\rm c}^{\rm Schw}-2)^2$, which implies
\begin{equation}
\beta_0^2=\frac{(x_{\rm c}^{\rm Schw})^3}{4-x_{\rm c}^{\rm Schw}}.
\end{equation}
Thus $b_{\rm crit}^{\rm Schw}=M\beta_0$ exactly reproduces Eq.~(\ref{eq:bcrit-vinf}). Positive $\alpha_{13}$ therefore increases $b_{\rm crit}$. If a probe is launched according to the Schwarzschild prediction,
\begin{equation}
b=b_{\rm crit}^{\rm Schw}(v_\infty)+\delta b ,
\end{equation}
its true distance from the Johannsen capture boundary is
\begin{equation}
\delta b_{\rm eff}
=\delta b-\Delta b_{\rm crit},
\qquad
\Delta b_{\rm crit}\simeq M\beta_1(v_\infty)\alpha_{13}.
\label{eq:delta-beff}
\end{equation}
For $M=10~M_\odot$ and $\alpha_{13}=10^{-5}$, this shift is $9.733$~cm for $v_\infty=0.2$ and $6.833$~cm for $v_\infty=0.3$. A trajectory that would be on the escaping side of the Schwarzschild threshold by $5$~cm can therefore be moved to the captured side by the deformation.

This is illustrated in Fig.~\ref{f-j}, computed with the numerical procedure described in Appendix~\ref{app:joh}. In both panels the impact parameter is fixed at $b=b_{\rm crit}^{\rm Schw}(M=10M_\odot,v_\infty)+5$~cm, while $\alpha_{13}$ is varied over $0$, $5\cdot10^{-8}$, $10^{-7}$, $5\cdot10^{-7}$, $10^{-6}$, $5\cdot10^{-6}$, $10^{-5}$, and $5\cdot10^{-5}$. For very small $\alpha_{13}$, the trajectory is close to the Schwarzschild one. As $\alpha_{13}$ increases, the true critical boundary moves outward, the effective offset in Eq.~(\ref{eq:delta-beff}) decreases, and the escaping probe accumulates a larger total azimuthal angle before returning to large radius. For example, for $v_\infty=0.2$, we find $\Delta\phi_{\rm tot}/(2\pi)\simeq 3.81$ for $\alpha_{13}=0$, $\Delta\phi_{\rm tot}/(2\pi)\simeq 3.85$ for $\alpha_{13}=10^{-6}$, and $\Delta\phi_{\rm tot}/(2\pi)\simeq 4.58$ for $\alpha_{13}=5\cdot10^{-6}$. The trajectory is then captured for $\alpha_{13}=10^{-5}$, so it no longer has an outgoing large-radius value of $\phi_{\rm out}$.

A swarm of probes with known asymptotic velocities and impact parameters brackets the capture boundary through capture/escape outcomes. For escaping near-critical probes, the continuous quantity that refines this bracket is the total azimuthal change $\Delta\phi_{\rm tot}=|\phi_{\rm i}-\phi_{\rm f}|$, with $\phi_{\rm i}$ and $\phi_{\rm f}$ evaluated at large radii on the ingoing and outgoing branches. The total angle contains a regular far-field contribution and the near-separatrix contribution computed in Section~\ref{s-zwm}; the latter is logarithmic in $|\delta b_{\rm eff}|$. Captured probes still contribute to the bracket through the capture outcome. For escaping probes near the deformed capture boundary,
\begin{equation}
\begin{split}
\Delta\phi_{\rm tot}\simeq
A_{\phi,{\rm eff}}\left[-\ln|\delta b_{\rm eff}|+C_{\phi,{\rm eff}}\right],
\\
A_{\phi,{\rm eff}}=2\pi S_{\rm eff},\qquad
S_{\rm eff}=S_{\rm Schw}+O(\alpha_{13}),
\\
S_{\rm Schw}=\frac{1}{2\pi}
\sqrt{\frac{r_{\rm c}^{\rm Schw}}{6M-r_{\rm c}^{\rm Schw}}},
\end{split}
\label{eq:dphitot-effective}
\end{equation}
where $C_{\phi,{\rm eff}}$ is the corresponding finite matching constant and $r_{\rm c}^{\rm Schw}$ is the Schwarzschild critical radius at the same $v_\infty$. The coefficient is fixed by the local expansion around the unstable orbit; for the linearized estimates below, the Schwarzschild coefficient is sufficient. The logarithmic near-separatrix behavior can sharpen the localization of the boundary relative to capture/escape bracketing alone, but it still requires accurate knowledge of the asymptotic data of the probes, namely $v_\infty$ and $b$. However, localizing the boundary at a single asymptotic velocity does not separate a metric deformation from a small shift in the black hole mass. This degeneracy is the focus of the next section.

\begin{figure*}[t]
\centering
\includegraphics[width=0.45\linewidth]{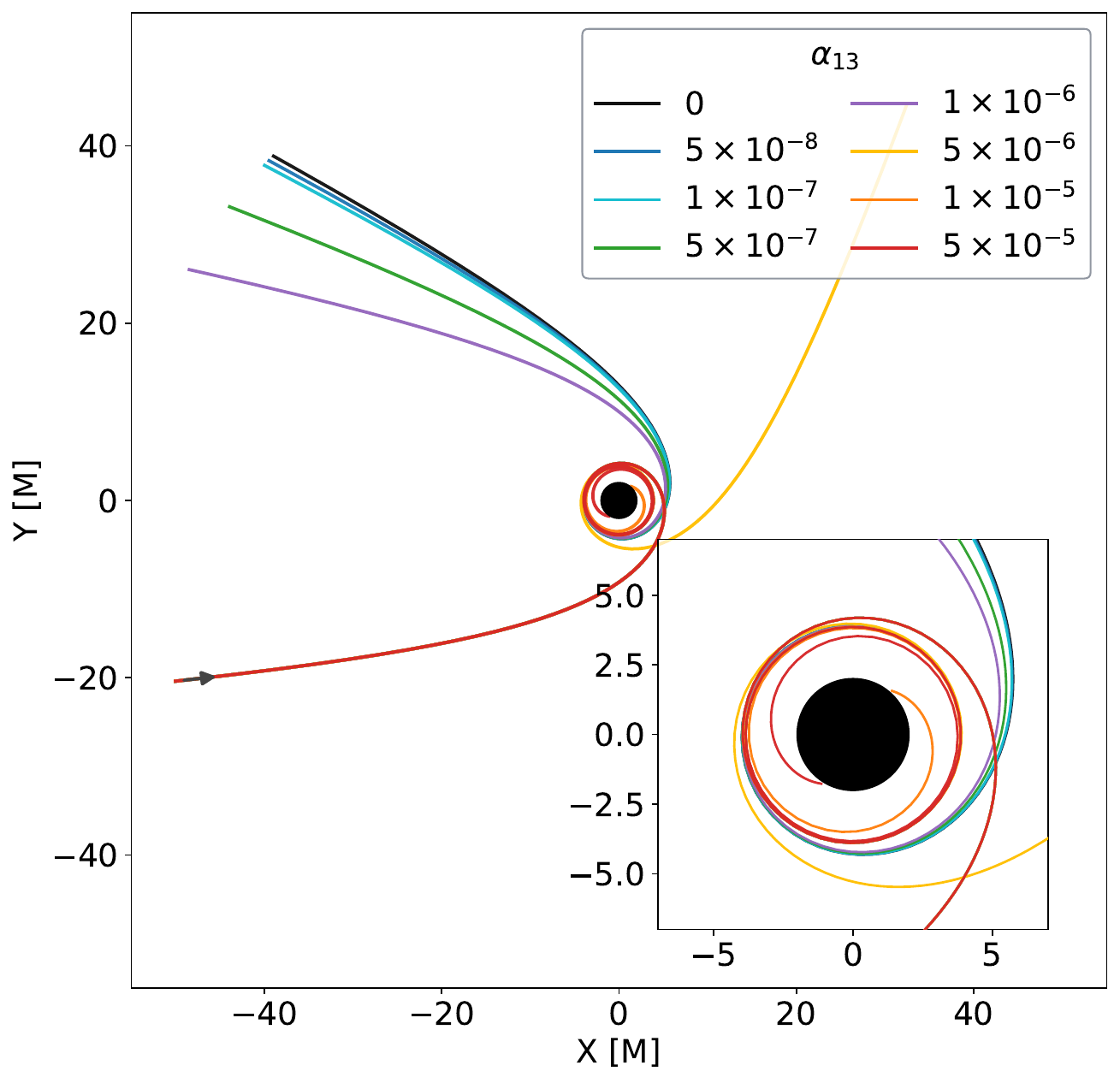}
\hspace{1.0cm}
\includegraphics[width=0.45\linewidth]{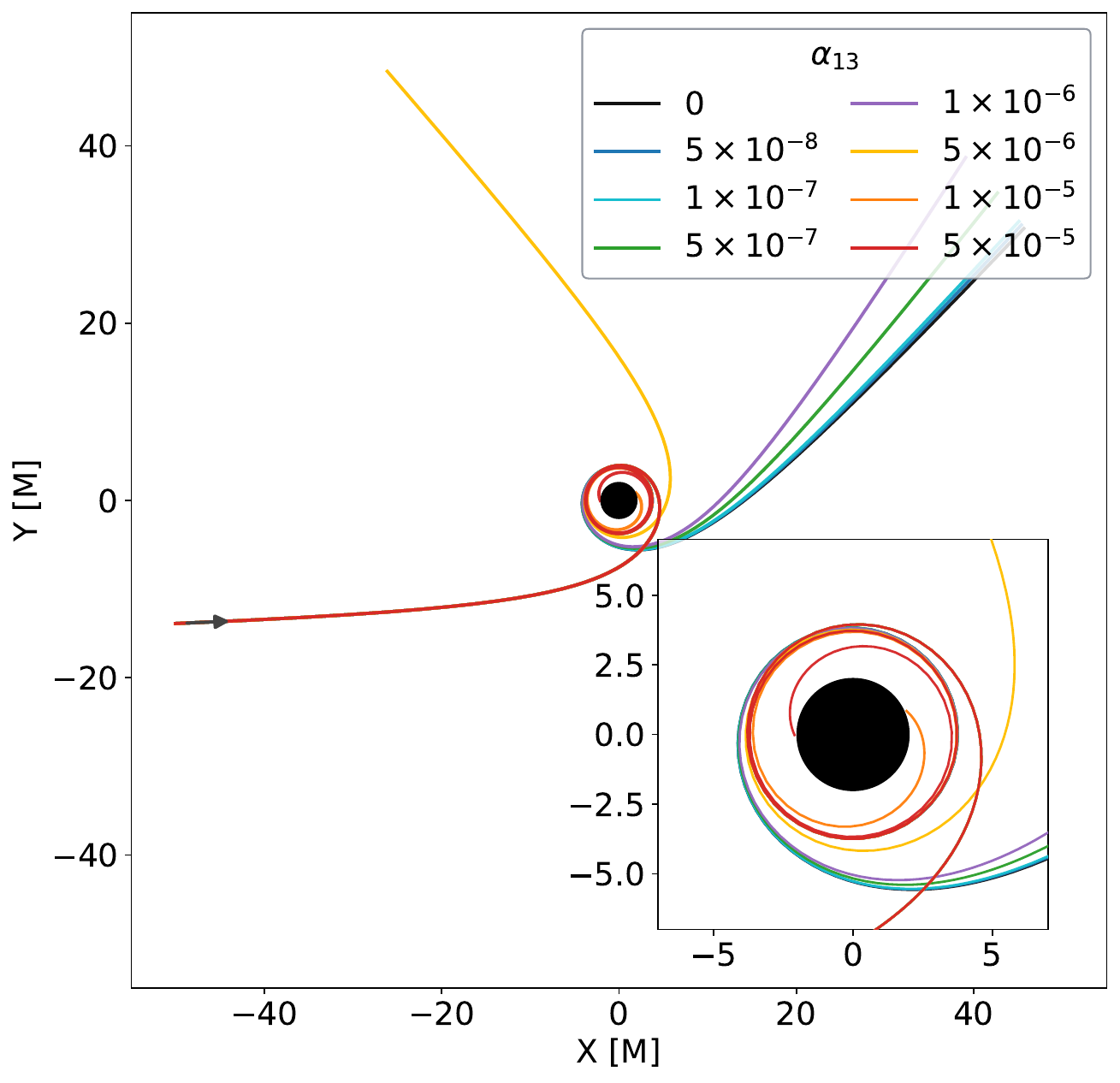}
\caption{Trajectories around Johannsen black holes with $M=10$~$M_\odot$. In each panel, the impact parameter is $b=b_{\rm crit}^{\rm Schw}(v_\infty)+5$~cm, where $b_{\rm crit}^{\rm Schw}$ is the Schwarzschild critical impact parameter for the same mass and asymptotic velocity. In the left panel, $v_\infty=0.2$, $b/M=20.38201551$, and $\ell=L/M=4.16046149$; in the right panel, $v_\infty=0.3$, $b/M=13.87901955$, and $\ell=4.36474972$. We show the trajectories for $\alpha_{13}=0$, $5\cdot10^{-8}$, $10^{-7}$, $5\cdot10^{-7}$, $10^{-6}$, $5\cdot10^{-6}$, $10^{-5}$, and $5\cdot10^{-5}$. The trajectories are integrated with DOP853 using relative tolerance $10^{-12}$, absolute tolerance $10^{-14}$, and maximum step $0.5~M$. The initial and termination conditions are given in Appendix~\ref{app:joh}. The black arrow indicates the propagation direction of the probes.
\label{f-j} }
\end{figure*}

\begin{figure}[t]
\centering
\includegraphics[width=0.9\linewidth]{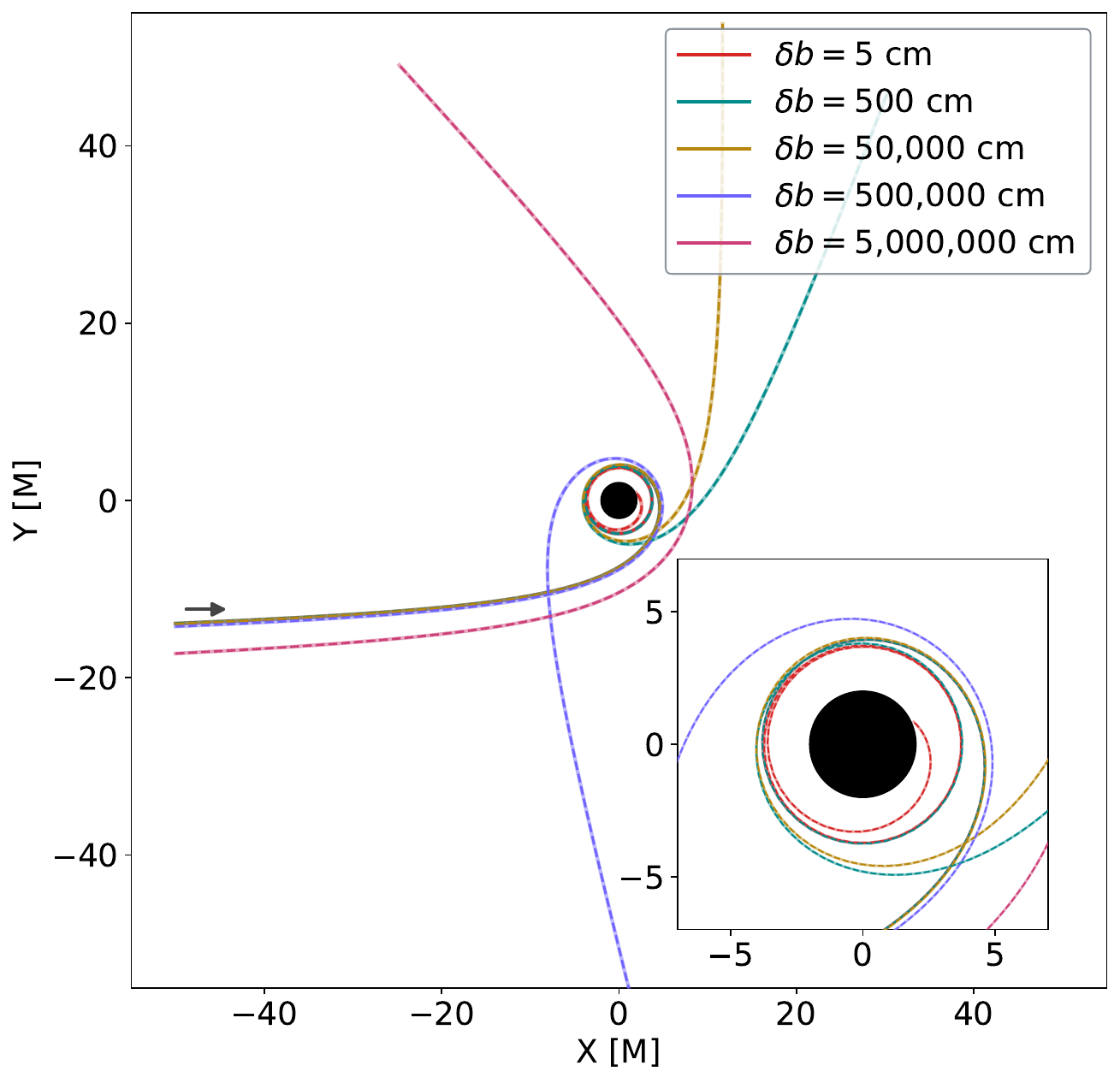}
\caption{Trajectories at $v_\infty=0.3$ around a Johannsen black hole with $M=10$~$M_\odot$ and $\alpha_{13}=10^{-5}$ (solid lines) and a Schwarzschild black hole with $M=10.00000333$~$M_\odot$ (dashed lines), whose physical critical impact parameter is matched to that of the Johannsen black hole. The colors correspond to $\delta b_{\rm label}=5$, 500, $5\cdot10^4$, $5\cdot10^5$, and $5\cdot10^6$~cm, where $\delta b_{\rm label}=b-b_{\rm crit}^{\rm Schw}$ and $b_{\rm crit}^{\rm Schw}$ is evaluated for a Schwarzschild black hole with $M=10$~$M_\odot$. Each paired solid and dashed trajectory has the same physical impact parameter $b$, with the specific angular momentum fixed by $L=\gamma_\infty v_\infty b$. The trajectories are integrated with DOP853 using relative tolerance $10^{-12}$, absolute tolerance $10^{-14}$, and maximum step $0.5M$ in the corresponding spacetime. The common physical initial conditions and the capture/escape criteria are given in Appendix~\ref{app:joh}. The black arrow indicates the propagation direction of the probes.
\label{f-b} }
\end{figure}

%%%%%%%%%%%%%%%%%%%%%%%%%%%%%%%%%%%%%%%%%%%%%%%%%%

\section{Mass-Geometry Degeneracy}\label{s-2}

\begin{figure}[t]
\centering
\includegraphics[width=0.9\linewidth]{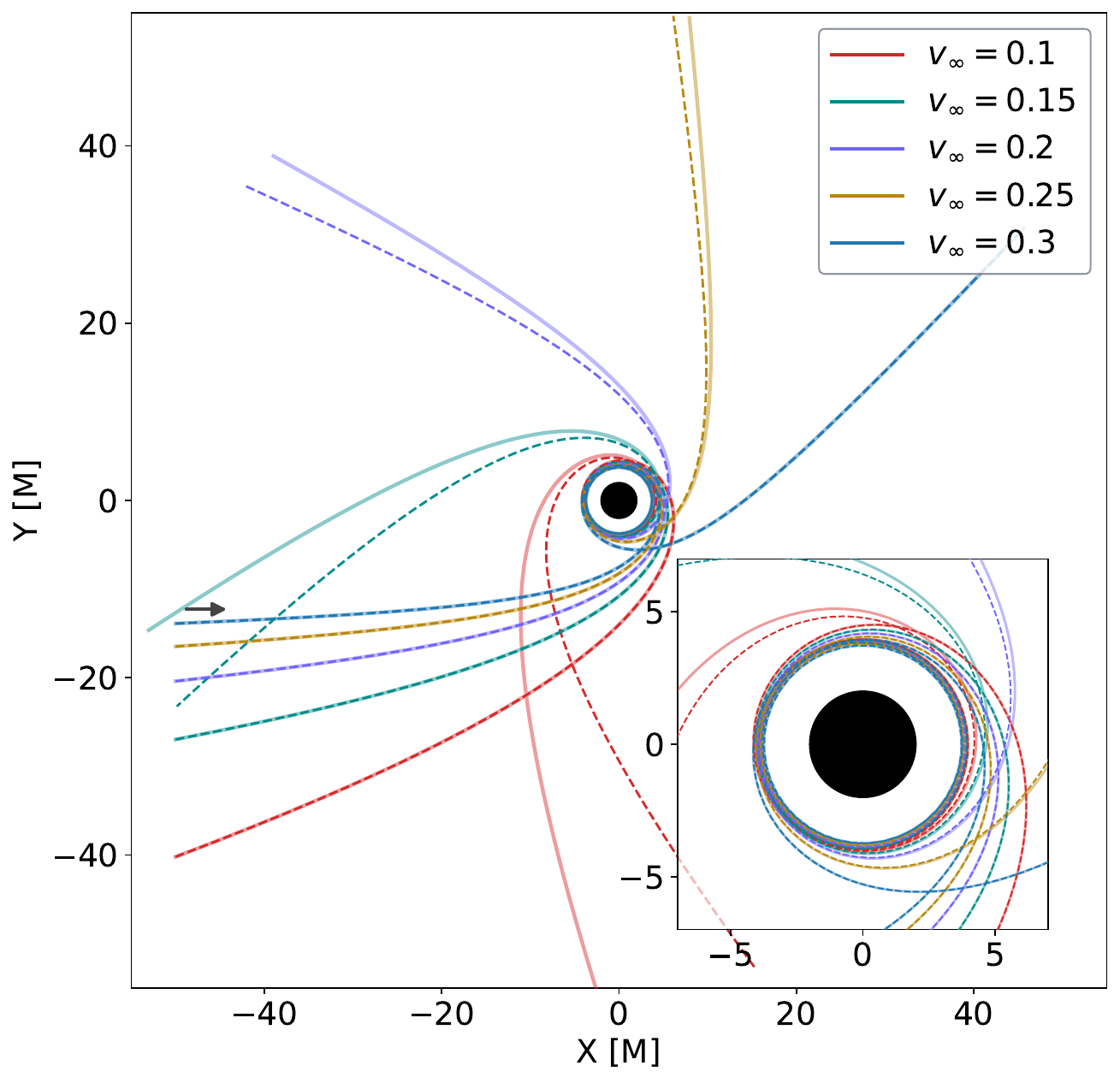}
\caption{Trajectories around a Johannsen black hole with $M = 10$~$M_\odot$ and $\alpha_{13} = 10^{-5}$ (solid lines) and a Schwarzschild black hole with the same critical impact parameter at $v_\infty = 0.3$ ($M = 10.00000333$~$M_\odot$, dashed lines) for $b-b_{\rm crit}^{\rm Joh}=5$~cm and different values of the asymptotic velocity $v_\infty$. The black arrow indicates the propagation direction of the probes.
\label{f-v}}
\end{figure}

A single-velocity near-critical flyby experiment can localize one quantity, the physical critical impact parameter $b_{\rm crit}(v_\infty)$, and therefore measures essentially one length scale. This creates a degeneracy between the black hole mass and metric deformations. Fig.~\ref{f-b} illustrates this degeneracy. The solid curves are for a Johannsen black hole with $M=10~M_\odot$ and $\alpha_{13}=10^{-5}$. The dashed curves are for a Schwarzschild black hole with a slightly larger mass, $M=10.0000033333~M_\odot$, chosen so that the physical critical impact parameter agrees with the Johannsen one at $v_\infty=0.3$. The labels in the figure are still referenced to the Schwarzschild critical impact parameter for $M=10~M_\odot$. At $v_\infty=0.3$, the matched Johannsen/mass-rescaled boundary is shifted outward by $6.833$~cm relative to this reference, so
\begin{equation}
\delta b_{\rm eff}
=\delta b_{\rm label}-6.833~{\rm cm}.
\label{eq:delta-beff-fb}
\end{equation}
The curve labeled $\delta b_{\rm label}=5$~cm is therefore actually on the captured side for both the Johannsen black hole and the mass-rescaled Schwarzschild black hole, with $\delta b_{\rm eff}=-1.833$~cm. 
The near-overlap of the two sets of trajectories is expected: at one velocity, most of the Johannsen shift in the critical impact parameter can be absorbed into a small change of the black hole mass. To see this explicitly, let us write the physical critical impact parameter as
\begin{equation}
b_{\rm crit}(v,\alpha_{13})
=M\left[\beta_0(v)+\beta_1(v)\alpha_{13}\right]
+O(\alpha_{13}^2),
\label{eq:Bcrit-joh}
\end{equation}
where $\beta_0$ and $\beta_1$ are defined in Eq.~(\ref{eq:kappa-joh}). A Schwarzschild black hole with mass $\tilde{M}$ has instead
\begin{equation}
b_{\rm crit}^{\rm Schw}(v;\tilde{M})=\tilde{M} \beta_0(v).
\end{equation}
If the two critical impact parameters are matched at a reference velocity $v_{\rm ref}$, we find
\begin{equation}
\tilde{M}=M\left[1+q(v_{\rm ref})\alpha_{13}\right],
\qquad
q(v)=\frac{\beta_1(v)}{\beta_0(v)}
=\frac{x_{\rm c}^{\rm Schw}-2}{(x_{\rm c}^{\rm Schw})^3}.
\label{eq:mass-match}
\end{equation}
For $v_{\rm ref}=0.3$, $M=10~M_\odot$, and $\alpha_{13}=10^{-5}$, Eq.~(\ref{eq:mass-match}) gives $\tilde{M}=10.0000033333~M_\odot$, as used in Figs.~\ref{f-b} and \ref{f-v}. The fractional mass change is only $3.3\cdot10^{-7}$. The degeneracy is therefore present only if the mass is treated as a free parameter at this level. If $M$ were known independently with better precision, for instance from far-field deflection or Keplerian dynamics, a single velocity measurement of $b_{\rm crit}$ would already constrain $\alpha_{13}$. In general, the problem should be regarded as a joint fit of $(M,\alpha_{13})$ to the localized function $b_{\rm crit}(v)$. This statement can be made explicit with two velocities. If $b_i=b_{\rm crit}(v_i,\alpha_{13})$ with $i=1,2$, then the ratio of the two measured critical impact parameters removes the overall mass scale at leading order,
\begin{equation}
\frac{b_1}{b_2}
=
\frac{\beta_0(v_1)}{\beta_0(v_2)}
\left[
1+\alpha_{13}
\left(
\frac{\beta_1(v_1)}{\beta_0(v_1)}
-\frac{\beta_1(v_2)}{\beta_0(v_2)}
\right)
\right]
+O(\alpha_{13}^2).
\label{eq:bcrit-ratio}
\end{equation}
Thus a single velocity fixes a normalization, while two or more velocities probe the shape of $b_{\rm crit}(v)$ and can separate a metric deformation from a pure rescaling of $M$.

After the mass is matched at one velocity, the remaining difference at another velocity is
\begin{equation}
\begin{split}
\Delta b_{\rm res}(v)
&=b_{\rm crit}^{\rm Joh}(v)
-b_{\rm crit}^{\rm Schw}(v;\tilde{M}) \\
&=M\alpha_{13}
\left[\beta_1(v)-q(v_{\rm ref})\beta_0(v)\right].
\end{split}
\label{eq:residual-bcrit}
\end{equation}
For the benchmark match at $v_{\rm ref}=0.3$, the raw Johannsen shifts and the residual shifts after the mass matching are summarized in Tab.~\ref{t-degeneracy}. The raw Johannsen shift at the matching velocity is $6.833$~cm, but the clean degeneracy-breaking residual at other velocities is only at the $0.3$-$1$~cm level in this example. Detecting that the capture boundary has moved is therefore easier than proving that the shift is caused by a metric deformation rather than by a small mass error. Let $\sigma_b$ denote the uncertainty with which the physical boundary $b_{\rm crit}(v)$ is localized at a second velocity, including the contribution from uncertainties in the probes' actual $v_\infty$ and $b$. The corresponding $1\sigma$ uncertainty on $\alpha_{13}$ after marginalizing over $M$ scales as
\begin{equation}
\sigma_{\alpha_{13}}(v|v_{\rm ref})
\simeq
\frac{\sigma_b}
{M\left|\beta_1(v)-q(v_{\rm ref})\beta_0(v)\right|}.
\label{eq:alpha-sigma}
\end{equation}
Sensitivity at the $10^{-5}$ level therefore requires this boundary-localization uncertainty to be below the residual scale in Tab.~\ref{t-degeneracy}. Since this scale is at or below the expected size of the small probes, the benchmark should be regarded as an idealized target rather than a demonstrated realistic precision.

After the conversion in Eq.~(\ref{eq:delta-beff-fb}), the Johannsen black hole and the mass-rescaled Schwarzschild black hole in Fig.~\ref{f-b} have the same physical distance from their shared capture boundary. The leading near-critical response is then the same logarithmic function of $|\delta b_{\rm eff}|$, Eq.~(\ref{eq:dphitot-effective}). The remaining differences are higher-order changes in the radial potential and in the physical time scale. At a single $v_\infty$, the data therefore constrain mainly the combination $M[\beta_0(v)+\beta_1(v)\alpha_{13}]$.

Figure~\ref{f-v} shows a velocity scan. We use the following conventions for the offsets. In Fig.~\ref{f-j}, the launch condition is measured from the fixed Schwarzschild boundary with $M=10~M_\odot$; in Fig.~\ref{f-b}, the plotted label is $\delta b_{\rm label}=b-b_{\rm crit}^{\rm Schw}$, while $\delta b_{\rm eff}$ is the distance from the actual boundary of the spacetime being compared. In Fig.~\ref{f-v}, the impact parameter is instead chosen separately at each velocity as
\begin{equation}
b(v_\infty)=b_{\rm crit}^{\rm Joh}(v_\infty)+5~{\rm cm},
\label{eq:fv-convention}
\end{equation}
where the superscript Joh denotes the Johannsen benchmark with $M=10~M_\odot$ and $\alpha_{13}=10^{-5}$. The same physical $b(v_\infty)$ is then used for the mass-matched Schwarzschild comparison. Thus the Johannsen benchmark is on the escaping side by $5$~cm at every velocity. This is not the same reference used in Fig.~\ref{f-b}, where the label $\delta b_{\rm label}=5$~cm corresponds to $\delta b_{\rm eff}=-1.833$~cm for both the Johannsen benchmark and the mass-matched Schwarzschild comparison and is therefore a captured trajectory.
Table~\ref{t-degeneracy} summarizes the boundary shifts for this same velocity-scan setup; the last column gives the effective distance of the mass-matched Schwarzschild trajectory from its own capture boundary in Fig.~\ref{f-v}.

\begin{table}[t]
\caption{Velocity dependence of the benchmark mass-geometry degeneracy. The second column gives the raw Johannsen shift of the critical impact parameter for $M=10~M_\odot$ and $\alpha_{13}=10^{-5}$. The third column gives the residual shift after matching the Johannsen and Schwarzschild critical impact parameters at $v_{\rm ref}=0.3$. The fourth column gives the effective offset of the mass-matched Schwarzschild comparison for the launch convention in Eq.~(\ref{eq:fv-convention}).}
\label{t-degeneracy}
\begin{ruledtabular}
\begin{tabular}{cccc}
$v_\infty$ &
$\Delta b_{\rm Joh}$ [cm] &
$\Delta b_{\rm res}$ [cm] &
$\delta b_{\rm eff}^{\rm match}$ [cm]\\
\hline
0.10 & 18.73 & $-1.058$ & 3.94\\
0.15 & 12.70 & $-0.572$ & 4.43\\
0.20 & 9.733 & $-0.302$ & 4.70\\
0.25 & 7.982 & $-0.125$ & 4.88\\
0.30 & 6.833 & 0 & 5.00
\end{tabular}
\end{ruledtabular}
\end{table}

Escaping probes give a second check of the mass-geometry degeneracy through the total proper time $\tau_{\rm tot}$ between two large-radius crossings, one on the ingoing branch and one on the outgoing branch. In the comparisons below, $\tau_{\rm tot}$ is evaluated at the same physical matching radius in the two spacetimes, so the endpoint convention does not introduce an artificial model difference. This radius is only the large-radius matching surface used to define the comparison, not a radius that must be identified experimentally. Near the separatrix, $\tau_{\rm tot}$ inherits a logarithmic contribution from the whirl region. For interpreting this logarithmic contribution, we use the whirl-region proper time only as a diagnostic. In dimensionless variables $x=r/M$, $\ell=L/M$, and $\hat\tau=\tau/M$, this diagnostic obeys
\begin{equation}
\hat\tau_{\rm whirl}
\simeq A_\tau\left[-\ln|\delta b_{\rm eff}|+C_\tau\right],
\qquad
A_\tau=\sqrt{\frac{2}{R_{xx}(x_{\rm c})}} ,
\label{eq:tau-whirl-log}
\end{equation}
where $R(x;\ell,\alpha_{13})=E^2h^2-f(1+\ell^2/x^2)$ is the dimensionless radial potential, and $R_{xx}(x_{\rm c})\equiv\partial^2_x R|_{x_{\rm c}}$ is evaluated at the critical unstable circular orbit of the spacetime under consideration. 
In Schwarzschild spacetime,
\begin{equation}
A_\tau^{\rm Schw}
=\sqrt{\frac{(x_{\rm c}^{\rm Schw})^3(x_{\rm c}^{\rm Schw}-3)}
{6-x_{\rm c}^{\rm Schw}}} .
\end{equation}
Matching $b_{\rm crit}$ at one velocity does not force the two metrics to have the same $M A_\tau$. For the benchmark Johannsen spacetime ($M=10~M_\odot$, $\alpha_{13}=10^{-5}$) and the mass-matched Schwarzschild comparison ($\tilde M=10.0000033333~M_\odot$ at $v_{\rm ref}=0.3$), the same-velocity timing residual is small: centimeter-scale offsets give total proper-time differences of only a few ns (Appendix~\ref{app:joh}). A velocity-scan comparison gives a larger residual. At each $v_\infty$, the Johannsen trajectory is launched at $b=b_{\rm crit}^{\rm Joh}(v_\infty)+5$~cm, the Schwarzschild comparison uses the same physical $b$, and $\tau_{\rm tot}$ is evaluated between the ingoing and outgoing crossings of the same physical matching radius $R_{\rm m}=50M_{\rm Joh}$. At $v_\infty=0.2$, this gives about $-14.8~\mu{\rm s}$. Increasing the common physical matching radius up to $10^4M_{\rm Joh}$ changes this value by only $0.017~\mu{\rm s}$, leaving the quoted value unchanged at this precision. The signal arises from the same residual boundary mismatch shown in Fig.~\ref{f-v}.

Thus only a velocity scan, or another independent family of initial data, can break the mass-geometry degeneracy.

%%%%%%%%%%%%%%%%%%%%%%%%%%%%%%%%%%%%%%%%%%%%%%%%%%

\section{Discussion and Conclusions}\label{s-c}

In the previous sections, we have studied the zoom-whirl motion in the Schwarzschild and a deformed-Schwarzschild spacetime. The aim of this section is to discuss how we can use these results to test the spacetime geometry of a non-rotating black hole with an interstellar mission of nanocrafts.

\begin{figure}[tbp]
\centering
\includegraphics[width=0.9\linewidth,trim=8cm 6cm 4cm 4.5cm,clip]{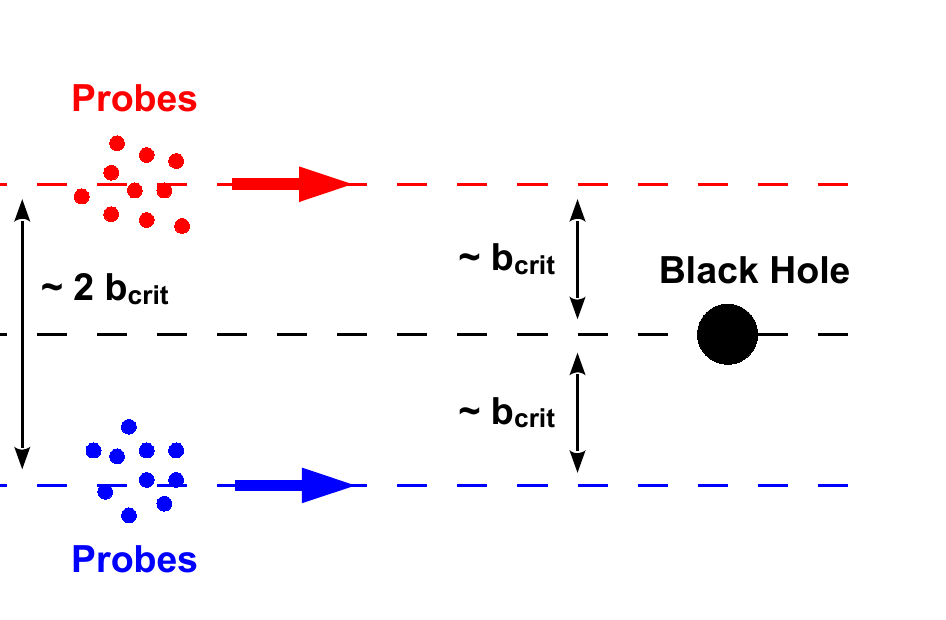}
%\vspace{0.0cm}
\caption{Two swarms of probes approaching the target black hole. The two swarms have impact parameters $b$ close to $b_{\rm crit}$ but are on opposite sides of the black hole. 
\label{f-m}}
\end{figure}

If the spacetime geometry is described by the Schwarzschild metric, there is only one parameter: the black hole mass $M$. We can measure the black hole mass by determining $b_{\rm crit}$ and $v_\infty$. There are several methods to measure $v_\infty$, ranging from the measurement of the Doppler shift of the electromagnetic signal from the nanocraft by a station on Earth to the comparison of the positions of three or more stars observed in the nanocraft’s comoving frame and in the Earth rest-frame~\cite{Zhu:2019trp}. For the measurement of $b_{\rm crit}$, we could, for instance, send two swarms of probes, as illustrated in the cartoon in Fig.~\ref{f-m}. These probes cannot be the nanocrafts with large light sails launched from Earth, because the light sails would be destroyed by tidal forces near the black hole. Instead, a nanocraft of a few grams in mass and a few meters in size could release a swarm of smaller (sub-gram and sub-centimeter) probes that can approach the target black hole with an impact parameter as close as possible to $b_{\rm crit}$. These small probes should be able to communicate with a nearby mothership (e.g., the nanocraft that released them) and the mothership should process and trasmit the data to Earth (the light sail could indeed be used as an antenna). The distance between the two swarms of small probes should be $\sim 2 b_{\rm crit}$ and can be measured with high precision via the exchange of electromagnetic signals between the two swarms far from the black hole. 
Every swarm can then measure $\delta b$ of one or more probes from measurements of $\Delta\phi_{\rm tot} = |\phi_{\rm i} - \phi_{\rm f}|$, where $\phi_{\rm i}$ is the value of $\phi$ far from the black hole when the probe is approaching the black hole and $\phi_{\rm f}$ is either the value of $\phi$ far from the black hole when the probe is escaping to infinity (for scattered orbits) or the value of $\phi$ when the probe is close to the black hole event horizon (for captured orbits).
$\Delta\phi_{\rm tot}$ can likely be inferred with high precision if the probes can receive a stable signal from a station on Earth and exchange electromagnetic signals among themselves. Potentially, we could measure $b_{\rm crit}$ with a precision of $\sim 1$~cm, which is roughly the size of these small probes. This would lead to a measurement of the black hole mass $M$ with a precision of order $10^{-7}$.

If we want to test the nature of the black hole, we must measure two parameters: the black hole mass $M$ and the deformation parameter (for example, $\alpha_{13}$). In this case, we could still use two swarms of small probes as in Fig.~\ref{f-m} and determine $b_{\rm crit}$, but this is alone is insufficient to measure $M$ and $\alpha_{13}$ simultaneously, as shown in Figure~\ref{f-b}. 
We need other measurements to break the mass-geometry degeneracy. In the previous section, we have explored the possibility of deploying swarms of small probes with different $v_\infty$ and that of tracking the proper time of these probes. However, these two solutions are not the only ones and alternative strategies also deserve to be considered.

The study presented in this work is focused on the measurements that could be performed by spacecraft in a possible future interstellar mission to test the nature of a black hole. The development of the necessary technology is certainly crucial for designing such a mission, but it is beyond the scope of the present study. We would also like to point out that this is a preliminary work with many simplifications. In particular, we have considered a non-rotating black hole in vacuum, neglected uncertainties in the black hole's position and velocity (which can presumably be measured with increasing precision by different swarms), and assumed that both the black hole and the probes move in a common plane. These idealizations should be relaxed in subsequent, more elaborate studies.

Turning to possible extensions, including black hole spin introduces an additional parameter that influences the nanocraft trajectories. A probe co-rotating with the black hole -- that is, with orbital angular momentum aligned with the spin -- will experience a slightly smaller critical impact parameter than in the Schwarzschild case, whereas the opposite holds for a counter-rotating probe. Given that we have already identified a degeneracy between the black hole mass and the deformation parameter $\alpha_{13}$, the spin will further affect the precise value of the critical impact parameter, thereby requiring additional independent observables to break the ensuing parameter degeneracy.

As for the presence of accreting matter surrounding the black hole, its mass is entirely negligible relative to the black hole mass and cannot appreciably modify the spacetime geometry (see, e.g., the discussion in Ref.~\cite{Bambi:2014koa}). However, the potential influence of this material on the probes' motion should not be dismissed a priori. The accreting medium may behave as a viscous fluid, possibly causing the nanocrafts to depart from purely geodesic motion. Consequently, while the critical impact parameter itself remains essentially unchanged, the equations of motion for the probes may need to be augmented to account for such dissipative effects in a more complete treatment.

In conclusion, considering the technical difficulties of decelerating the nanocrafts and placing them on specific orbits as close as possible to the black hole (Paper~I), the flyby experiment presented in this work may offer a promising approach to testing the nature of the black hole. Both strategies present many challenges, and future studies will have to assess the actual feasibility of the two approaches.

%%%%%%%%%%%%%%%%%%%%%%%%%%%%%%%%%%%%%%%%%%%%%%%%%%

\section*{Acknowledgements}
%\section*{Funding}

This work was supported by the National Natural Science Foundation of China (NSFC), Grant No.~W2531002.

%\section*{Data Availability Statement}

%\section*{Code Availability Statement}

%%%%%%%%%%%%%%%%%%%%%%%%%%%%%%%%%%%%%%%%%%%%%%%%%%

\appendix

\section{Numerical procedure for the Schwarzschild flybys}\label{app:numerics}

In this appendix, we summarize the numerical procedure used for Figs.~\ref{f-nw1}, \ref{f-nw2}, and \ref{f-sch}. For every choice of $(v_\infty,b)$, we first compute
\begin{equation}
E=\gamma_\infty=\frac{1}{\sqrt{1-v_\infty^2}},
\qquad
L=\gamma_\infty v_\infty b .
\end{equation}
We then integrate the equatorial timelike geodesic equations in the Schwarzschild spacetime as a first-order system in the probe proper time,
\begin{equation}
\dot r=u,\qquad
\dot u=-\frac{M}{r^2}+\frac{L^2}{r^3}
-\frac{3ML^2}{r^4},\qquad
\dot\phi=\frac{L}{r^2}.
\label{eq:numerical-odes}
\end{equation}
These equations are equivalent to Eq.~(\ref{eq:radial-sch}) and the conservation of $L$.

For the one-dimensional scans in Figs.~\ref{f-nw1} and \ref{f-nw2}, the finite matching radius is $R_{\rm m}=50M$. The integration starts at $r_0=R_{\rm m}$ on the ingoing branch, with
\begin{equation}
u_0=-\left[
E^2-\left(1-\frac{2M}{r_0}\right)
\left(1+\frac{L^2}{r_0^2}\right)
\right]^{1/2}.
\end{equation}
The escape event is activated only on the outgoing branch, so the initial point at $r=R_{\rm m}$ is not counted as an escape. For escaping trajectories, we record the unwrapped azimuthal angle at the ingoing and outgoing crossings of $R_{\rm m}$ and compute
\begin{equation}
\Delta\phi_{\rm tot}
=|\phi_{\rm out}(R_{\rm m})-\phi_{\rm in}(R_{\rm m})| .
\end{equation}
Captured trajectories have no outgoing large-radius endpoint and are used only to bracket the capture boundary. The finite value of $R_{\rm m}$ affects the regular additive part of $\Delta\phi_{\rm tot}$, not the local logarithmic coefficient. The initial value of $\phi$ only fixes the orientation of the trajectory in the equatorial plane. For the trajectories shown in Fig.~\ref{f-sch}, we instead choose the initial Cartesian position $(X_0,Y_0)=(-50M,-b)$, so that $r_0=(X_0^2+Y_0^2)^{1/2}$ and $\phi_0$ is the polar angle of $(X_0,Y_0)$ in the correct quadrant, equivalently $\phi_0=\operatorname{atan2}(Y_0,X_0)$ if one uses the two-argument inverse tangent, again with the ingoing radial branch. For all integrations, a trajectory is classified as capture when the probe reaches $r=1.05\,r_{\rm H}=2.1M$. We used an eighth-order explicit Runge-Kutta method (DOP853) with relative tolerance $10^{-12}$, absolute tolerance $10^{-14}$, maximum step $0.5M$, and maximum integration time $10^5M$.

For diagnostic purposes, we also determine the radius $r_{\rm u}$ of the maximum of $V_{\rm eff}(r;L)$. The numerical whirl phase is defined as the part of the trajectory satisfying $r<1.3r_{\rm u}$. If $\phi_{\rm in}$ and $\phi_{\rm out}$ are the values of $\phi$ at the first and last points in this region, the numerical whirl number is
\begin{equation}
n_{\rm whirl}=
\frac{|\phi_{\rm out}-\phi_{\rm in}|}{2\pi}.
\label{eq:nwhirl-numerical}
\end{equation}
Changing the numerical factor 1.3 in the definition of the whirl region changes only the finite constant in the near-critical logarithmic fit, not the logarithmic slope. This diagnostic is not used as the experimental observable in Figs.~\ref{f-nw1} and \ref{f-nw2}; it only checks the local logarithmic behavior. Similar logarithmic behavior is found in black hole encounters near the immediate-merger threshold~\cite{Pretorius:2007jn}.

The constant $\kappa$ in Eq.~(\ref{eq:near-critical-radial}) is defined by $R(r_{\rm c};\lambda)=-\kappa\Delta+O(\Delta^2)$. At fixed $v_\infty$ and with $\Delta_b= (b-b_{\rm crit})/M$, one has
\begin{equation}
\kappa_b=
-M\left(\frac{\partial R}{\partial b}\right)_{\rm c}
=
M \gamma_\infty v_\infty\,
\frac{2L_{\rm c}}{r_{\rm c}^2}
\left(1-\frac{2M}{r_{\rm c}}\right).
\end{equation}
At fixed $b$ and with $\Delta_v= v_\infty-v_{\rm crit}$, one finds
\begin{equation}
\kappa_v=
-\left(\frac{\partial R}{\partial v_\infty}\right)_{\rm c}
=
\frac{2L_{\rm c}}{r_{\rm c}^2}
\left(1-\frac{2M}{r_{\rm c}}\right)
bE_{\rm c}^3
-
2v_{\rm crit}E_{\rm c}^4 ,
\end{equation}
where $E_{\rm c}=\gamma_\infty(v_{\rm crit})$ and $L_{\rm c}=E_{\rm c}v_{\rm crit}b$. Both coefficients are positive for the families considered here, consistent with the convention that $\Delta>0$ denotes the scattered side.

The dashed lines in Figs.~\ref{f-nw1} and \ref{f-nw2} use the analytical slope
\begin{equation}
S=\frac{1}{2\pi}
\sqrt{\frac{r_{\rm c}}{6M-r_{\rm c}}}.
\end{equation}
Only the additive constants are fitted to the numerical data. For a given escaping branch, let $Y_i=\Delta\phi_{{\rm tot},i}/(2\pi)$ be the plotted value of the $i$-th numerical trajectory, and let $\Delta_i$ be the horizontal variable used for that trajectory in the corresponding plot, namely $\Delta_b$ in Fig.~\ref{f-nw1} or $\Delta_v$ in Fig.~\ref{f-nw2}. Using the $N$ points closest to the critical line, we set
\begin{equation}
C_\phi =\frac{1}{N}\sum_{i=1}^N
\left(\frac{Y_i}{S}+\ln |\Delta_i|\right),
\end{equation}
with $N=8$ in the figures shown here. If the plotted horizontal variable is expressed in centimeters rather than in units of $M$, this only shifts $C_\phi$ and leaves the slope unchanged.

%%%%%%%%%%%%%%%%%%%%%%%%%%%%%%%%%%%%%%%%%%%%%%%%%%

\section{Numerical procedure for the Johannsen flybys}\label{app:joh}

In this appendix, we collect the analytic and numerical details used in Sections~\ref{s-1} and \ref{s-2} for the Johannsen spacetime of Eq.~(\ref{eq-j}) and for the mass-matched comparison.

We first derive the first-order shift of the critical impact parameter. We write the right-hand side of the turning-point relation, Eq.~(\ref{eq:turning-joh}), as $V_{\rm eff}(x,\ell,\alpha_{13})$, with the dimensionless variables $x=r/M$ and $\ell=L/M$,
\begin{equation}
V_{\rm eff}=
\frac{\left(1-2/x\right)\left(1+\ell^2/x^2\right)}
{\left(1+\alpha_{13}/x^3\right)^2}.
\end{equation}
Define
\begin{equation}
F(x,\ell,\alpha_{13})\equiv E^2-V_{\rm eff}(x,\ell,\alpha_{13}) .
\end{equation}
The critical trajectory satisfies $F=0$ and $\partial_x F=0$ at the critical radius $x=x_{\rm c}$. We write
\begin{equation}
\begin{split}
x_{\rm c}&=x_{\rm c}^{\rm Schw}+\alpha_{13}x_1+O(\alpha_{13}^2), \\
\ell_{\rm c}&=\ell_{\rm c}^{\rm Schw}+\alpha_{13}\ell_1+O(\alpha_{13}^2),
\end{split}
\end{equation}
where $x_{\rm c}^{\rm Schw}$ and $\ell_{\rm c}^{\rm Schw}$ are the critical values of $x$ and $\ell$ for the Schwarzschild case $\alpha_{13} = 0$. Denoting $F_i\equiv \partial_i F$ and $F_{ij}\equiv \partial_i \partial_j F$, the linearized criticality conditions at fixed $E$ are
\begin{align}
0&=F_x x_1+F_\ell \ell_1+F_\alpha, \\
0&=F_{xx}x_1+F_{x\ell}\ell_1+F_{x\alpha},
\end{align}
where all derivatives are evaluated at $\alpha_{13}=0$, $x=x_{\rm c}^{\rm Schw}$, and $\ell=\ell_{\rm c}^{\rm Schw}$. Since $F_x=0$ at the critical point, the first equation directly gives
\begin{equation}
\ell_1\equiv\frac{d\ell_{\rm c}}{d\alpha_{13}}
=-\frac{F_\alpha}{F_\ell}
=-\frac{V_\alpha}{V_\ell}.
\end{equation}
The second equation determines $x_1$ after $\ell_1$ is known and is not needed for the critical-angular-momentum shift. At the Schwarzschild critical point, using $F=0$,
\be
V_\alpha\big|_{\rm c}
&=&-\frac{2}{\left(x_{\rm c}^{\rm Schw}\right)^3}\left(1-\frac{2}{x_{\rm c}^{\rm Schw}}\right)
\left[1+\left(\frac{\ell_{\rm c}^{\rm Schw}}{x_{\rm c}^{\rm Schw}}\right)^2\right] \nonumber\\
&=&-\frac{2E^2}{\left( x_{\rm c}^{\rm Schw} \right)^3}, \\
V_\ell\big|_{\rm c}
&=&\left(1-\frac{2}{x_{\rm c}^{\rm Schw}}\right)\frac{2\ell_{\rm c}^{\rm Schw}}{\left( x_{\rm c}^{\rm Schw} \right)^2}
=\frac{2\ell_{\rm c}^{\rm Schw}(x_{\rm c}^{\rm Schw}-2)}{\left( x_{\rm c}^{\rm Schw} \right)^3}. \nonumber\\
\ee
Therefore,
\begin{equation}
\ell_1= \frac{E^2}{\ell_{\rm c}^{\rm Schw}(x_{\rm c}^{\rm Schw}-2)}.
\end{equation}
At the Schwarzschild critical orbit,
\be
E&=&E_{\rm c}= \frac{x_{\rm c}^{\rm Schw} -2}{\sqrt{x_{\rm c}^{\rm Schw}(x_{\rm c}^{\rm Schw} - 3)}} \, , \\
\ell_{\rm c}^{\rm Schw} &=& \frac{x_{\rm c}^{\rm Schw}}{\sqrt{x_{\rm c}^{\rm Schw}-3}} \, ,
\ee
from Eq.~(\ref{eq:circular-energy-angular}), so
\begin{equation}
\ell_1= \frac{x_{\rm c}^{\rm Schw}-2}{\left( x_{\rm c}^{\rm Schw} \right)^2\sqrt{x_{\rm c}^{\rm Schw}-3}}.
\end{equation}
Since $b/M=\ell/(\gamma_\infty v_\infty)$ and $\gamma_\infty=E_{\rm c}$, the first-order coefficient in Eq.~(\ref{eq:kappa-joh}) is
\begin{equation}
\beta_1=\frac{\ell_1}{\gamma_\infty v_\infty}
=\frac{1}{v_\infty \left( x_{\rm c}^{\rm Schw} \right)^{3/2}},
\end{equation}
and the mass-matching coefficient in Eq.~(\ref{eq:mass-match}) becomes
\begin{equation}
q(v)=\frac{\beta_1(v)}{\beta_0(v)}=\frac{x_{\rm c}^{\rm Schw}-2}{(x_{\rm c}^{\rm Schw})^3}.
\end{equation}
We verified numerically, at $v_\infty=0.1$, 0.2, 0.25, and 0.3, that the full nonlinear shift $b_{\rm crit}(v_\infty,10^{-5})-b_{\rm crit}(v_\infty,0)$ agrees with the linearized expression $M\beta_1\alpha_{13}$ to relative accuracy $\sim 10^{-6}$, so the $O(\alpha_{13}^2)$ terms are negligible at the benchmark deformation.

For a given $v_\infty$ (hence $E=\gamma_\infty$) and $\alpha_{13}$, the critical impact parameter is obtained by a one-dimensional root find. For each trial value of $b/M$, we evaluate the maximum of $V_{\rm eff}(x,\ell,\alpha_{13})$ over $x$ and bisect the bracket separating $E^2>V_{\rm eff}(x_{\rm u})$ from $E^2<V_{\rm eff}(x_{\rm u})$, where $x_{\rm u}$ is the barrier-top radius. The bisection is terminated when the bracket width in $b/M$ is below $10^{-12}$, and $b_{\rm crit}=M\ell/(\gamma_\infty v_\infty)$ at the boundary. The matched Schwarzschild mass is $\tilde{M}=M[1+q(v_{\rm ref})\alpha_{13}]$; for $M=10~M_\odot$, $\alpha_{13}=10^{-5}$, and $v_{\rm ref}=0.3$ this gives $\tilde{M}=10.0000033333~M_\odot$, used in Figs.~\ref{f-b} and \ref{f-v}.

The equatorial timelike geodesics in the spacetime of Eq.~(\ref{eq-j}) are integrated as the first-order system
\begin{equation}
\dot r=u,\;\,
\dot u=-\frac{M}{r^2}+\frac{L^2}{r^3}-\frac{3ML^2}{r^4}
-\frac{3\alpha_{13}E^2M^3 h}{r^4},\;\,
\dot\phi=\frac{L}{r^2},
\end{equation}
with $h=1+\alpha_{13}M^3/r^3$, which reduces to Eq.~(\ref{eq:numerical-odes}) for $\alpha_{13}=0$. We use the eighth-order DOP853 integrator with relative tolerance $10^{-12}$, absolute tolerance $10^{-14}$, maximum step $0.5M$, and maximum integration time $10^5M$. The horizon stays at $r_{\rm H}=2M$ because the deformation $h$ does not affect $g_{rr}$. For the trajectory plots, capture is recorded at $r=1.05\,r_{\rm H}=2.1M$ and escape is recorded at the outer stop radius used to terminate the integration, $r=55M$ in the reference-mass units. This stop radius is not used as the observational endpoint for the angular and timing observables below. For the single-mass trajectory scan in Fig.~\ref{f-j}, the initial conditions are again set at $(X_0,Y_0)=(-50M,-b)$. In Figs.~\ref{f-b} and \ref{f-v}, all plotted lengths are in units of the Johannsen-benchmark mass $M_{\rm Joh}=10M_\odot$, and the Johannsen benchmark and the mass-matched Schwarzschild comparison are launched from the same physical starting point on the ingoing branch
\begin{equation}
(X_0,Y_0)=(-50M_{\rm Joh},-b),
\end{equation}
where $b$ is the physical impact parameter. 
For the large-radius observables, we use escaping trajectories and evaluate both $\Delta\phi_{\rm tot}$ and $\tau_{\rm tot}$ between the ingoing and outgoing crossings of the same physical matching radius $R_{\rm m}=50M_{\rm Joh}$, so that the angular and timing residuals are not contaminated by different endpoint choices when the masses differ. This radius is a large-radius matching convention rather than an experimentally specified surface. The outgoing crossing of $R_{\rm m}$ is located as an integration event.
In Fig.~\ref{f-j}, $b=b_{\rm crit}^{\rm Schw}(M=10M_\odot,v_\infty)+5$~cm is held fixed while $\alpha_{13}$ is scanned; in Fig.~\ref{f-b}, the Johannsen benchmark ($M=10M_\odot$, $\alpha_{13}=10^{-5}$) and the mass-matched Schwarzschild comparison ($\tilde{M}$) are launched at the same physical $b$, with the plotted label $\delta b_{\rm label}$ referenced to the $M=10M_\odot$ Schwarzschild boundary; in Fig.~\ref{f-v}, the launch is $b=b_{\rm crit}^{\rm Joh}(v_\infty)+5$~cm at each velocity. The whirl number is computed as in Eq.~(\ref{eq:nwhirl-numerical}), with the barrier-top radius $r_{\rm u}$ evaluated in the Johannsen potential, and is used only as a diagnostic of the local logarithmic behavior.

Finally, the elapsed proper time is $\hat\tau=\tau/M=\int dx/\sqrt{R}$, with $R=E^2h^2-f(1+\ell^2/x^2)$ and $f=1-2/x$. Expanding $R$ about the critical radius $x_{\rm c}$ as $R\simeq C_b\,\delta b_{\rm eff}+\tfrac{1}{2}R_{xx}(x_{\rm c}) (x-x_{\rm c})^2$, where $C_b$ is a finite coefficient for the chosen one-parameter launch family, gives the logarithmic law in Eq.~(\ref{eq:tau-whirl-log}), with $A_\tau=\sqrt{2/R_{xx}(x_{\rm c})}$; in the Schwarzschild limit $R_{xx}(x_{\rm c}^{\rm Schw})=2(6-x_{\rm c}^{\rm Schw})/[(x_{\rm c}^{\rm Schw})^3(x_{\rm c}^{\rm Schw}-3)]$ and $A_\tau^{\rm Schw}=\sqrt{(x_{\rm c}^{\rm Schw})^3(x_{\rm c}^{\rm Schw}-3)/(6-x_{\rm c}^{\rm Schw})}$. The whirl-region proper time is accumulated over the same operational region $r<1.3\,r_{\rm u}$ used for $n_{\rm whirl}$ and is used only as a diagnostic of the logarithmic behavior; the whirl endpoints are located by solving for the crossings of $r=1.3\,r_{\rm u}$. The experimentally relevant quantity is the total proper time $\tau_{\rm tot}$ of escaping trajectories between ingoing and outgoing crossings of the same large radius, obtained from the same event-based integration. After $b_{\rm crit}$ is matched at one velocity, the same-velocity proper-time difference between the Johannsen benchmark and the mass-matched Schwarzschild comparison is at the nanosecond level for centimeter-scale offsets. For the velocity-scan comparison quoted in Section~\ref{s-2}, the mass is matched at $v_{\rm ref}=0.3$, the Johannsen benchmark is launched at $b=b_{\rm crit}^{\rm Joh}(v_\infty)+5$~cm, the Schwarzschild comparison uses the same physical $b$, and at $v_\infty=0.2$ the resulting same-radius residual is about $-14.8~\mu$s.

%%%%%%%%%%%%%%%%%%%%%%%%%%%%%%%%%%%%%%%%%%%%%%%%%%


\begin{thebibliography}{99}

\bibitem{Orion}
G.~Dyson,
{\it Project Orion: The True Story of the Atomic Spaceship}
(Henry Holt and Co, 2002),
ISBN 978-0805059854.

\bibitem{Daedalus}
K.~F.~Long and P.~R.~Galea,
{\it Project Daedalus: Demonstrating the Engineering Feasibility of Interstellar Travel}
(British Interplanetary Society, 2015),
ISBN 978-0950659701.

\bibitem{Marx}
G.~Marx,
{\it Interstellar Vehicle Propelled By Terrestrial Laser Beam},
Nature \textbf{211}, 22-23 (1966),
\href{https://doi.org/10.1038/211022a0}{https://doi.org/10.1038/211022a0}

\bibitem{Redding}
J.~L.~Redding,
{\it Interstellar Vehicle propelled by Terrestrial Laser Beam},
Nature \textbf{213}, 588-589 (1967),
\href{https://doi.org/10.1038/213588a0}{https://doi.org/10.1038/213588a0}

\bibitem{Lubin16}
P.~Lubin,
{\it A Roadmap to Interstellar Flight},
Journal of the British Interplanetary Society \textbf{69}, 40-72 (2016)
[arXiv:1604.01356 [astro-ph.EP]].

\bibitem{Lubin22}
P.~Lubin,
{\it The Path to Transformational Space Exploration}
(World Scientific Publishing Company, 2022),
ISBN 978-981-12-4903-7, 978-981-12-4828-3,
\href{https://doi.org/10.1142/11918}{https://doi.org/10.1142/11918}

\bibitem{Parkin18}
K.~L.~G.~Parkin,
{\it The Breakthrough Starshot system model},
Acta Astronautica \textbf{152}, 370-384 (2018),
\href{https://doi.org/10.1016/j.actaastro.2018.08.035}{https://doi.org/10.1016/j.actaastro.2018.08.035}
[arXiv:1805.01306 [astro-ph.IM]].

\bibitem{Kuhlmey25}
J.~Y.~Lin, C.~M.~de~Sterke, O.~Ilic and B.~T.~Kuhlmey,
{\it Lightsails for Interstellar Travel: Photonics for Propulsion, Thermal Management and Stability},
ACS Photonics \textbf{12}, 4818-4850 (2025),
\href{https://doi.org/10.1021/acsphotonics.5c00450}{https://doi.org/10.1021/acsphotonics.5c00450}
[arXiv:2502.17828 [astro-ph.IM]].

\bibitem{Eubanks2026}
T.~M.~Eubanks, J.~Schneider, B.~Bills, et al.,
{\it Science from the In Situ Exploration of the Proxima Centauri System},
\href{https://doi.org/10.48550/arXiv.2604.20182}{https://doi.org/10.48550/arXiv.2604.20182}
[arXiv:2604.20182 [astro-ph.IM]].

\bibitem{Bambi:2025kcr}
C.~Bambi,
{\it An interstellar mission to test astrophysical black holes},
iScience \textbf{28}, 113142 (2025), 
\href{https://doi.org/10.1016/j.isci.2025.113142}{https://doi.org/10.1016/j.isci.2025.113142}
[arXiv:2504.14576 [gr-qc]].

\bibitem{Bambi:2025hjn}
C.~Bambi,
{\it An interstellar mission to the closest black hole?},
\href{https://doi.org/10.48550/arXiv.2509.11222}{https://doi.org/10.48550/arXiv.2509.11222}
[arXiv:2509.11222 [gr-qc]].

\bibitem{Bambi:2026dwc}
C.~Bambi,
{\it A Space Mission to Earth's Nearest Black Hole: Reality or Science Fiction?},
\href{https://doi.org/10.48550/arXiv.2607.10982}{https://doi.org/10.48550/arXiv.2607.10982}
[arXiv:2607.10982 [gr-qc]].

\bibitem{Bambi:2017khi}
C.~Bambi,
{\it Black Holes: A Laboratory for Testing Strong Gravity}
(Springer Singapore, 2017),
ISBN 978-981-10-4523-3, 978-981-13-5158-7, 978-981-10-4524-0,
\href{https://doi.org/10.1007/978-981-10-4524-0}{https://doi.org/10.1007/978-981-10-4524-0}.

\bibitem{Bambi:2015kza}
C.~Bambi,
{\it Testing black hole candidates with electromagnetic radiation},
Rev. Mod. Phys. \textbf{89}, 025001 (2017),
\href{https://doi.org/10.1103/RevModPhys.89.025001}{https://doi.org/10.1103/RevModPhys.89.025001}
[arXiv:1509.03884 [gr-qc]].

\bibitem{Yagi:2016jml}
K.~Yagi and L.~C.~Stein,
{\it Black Hole Based Tests of General Relativity},
Class. Quant. Grav. \textbf{33}, 054001 (2016),
\href{https://doi.org/10.1088/0264-9381/33/5/054001}{https://doi.org/10.1088/0264-9381/33/5/054001}
[arXiv:1602.02413 [gr-qc]].

\bibitem{Murchikova:2025oio}
L.~Murchikova and K.~C.~Sahu,
{\it Observability of Isolated Stellar-mass Black Holes},
Astrophys. J. Lett. \textbf{988}, L12 (2025),
\href{https://doi.org/10.3847/2041-8213/ade7f8}{https://doi.org/10.3847/2041-8213/ade7f8}
[arXiv:2506.20711 [astro-ph.GA]].

\bibitem{Nosirov:2026fjo}
A.~Nosirov, C.~Bambi, L.~Gao, J.~de Bruijne, J.~Jiang, A.~Santangelo and F.~G.~Xie,
{\it Searching for Isolated Black Hole Candidates within 15 pc of the Solar System in Gaia DR3},
Astrophys. J. \textbf{1004}, 21 (2026),
\href{https://doi.org/10.3847/1538-4357/ae6805}{https://doi.org/10.3847/1538-4357/ae6805}
[arXiv:2601.14499 [astro-ph.HE]].

\bibitem{Nosirov2}
A.~Nosirov, C.~Bambi, L.~Gao, {\it et al.},
in preparation.

\bibitem{Gao:2026jpl}
L.~Gao, C.~Bambi, Y.~Fan, T.~Mirzaev, A.~Nosirov and A.~Santangelo,
{\it Testing Black Holes with Interstellar Missions: I. Orbiting Probes},
\href{https://doi.org/10.48550/arXiv.2605.19176}{https://doi.org/10.48550/arXiv.2605.19176}
[arXiv:2605.19176 [gr-qc]].

\bibitem{Glampedakis:2002ya}
K.~Glampedakis and D.~Kennefick,
{\it Zoom and whirl: Eccentric equatorial orbits around spinning black holes and their evolution under gravitational radiation reaction},
Phys. Rev. D \textbf{66}, 044002 (2002),
\href{https://doi.org/10.1103/PhysRevD.66.044002}{https://doi.org/10.1103/PhysRevD.66.044002}
[arXiv:gr-qc/0203086 [gr-qc]].

\bibitem{Johannsen:2013szh}
T.~Johannsen,
{\it Regular Black Hole Metric with Three Constants of Motion},
Phys. Rev. D \textbf{88}, 044002 (2013),
\href{https://doi.org/10.1103/PhysRevD.88.044002}{https://doi.org/10.1103/PhysRevD.88.044002}
[arXiv:1501.02809 [gr-qc]].

\bibitem{Tripathi:2020yts}
A.~Tripathi, Y.~Zhang, A.~B.~Abdikamalov, D.~Ayzenberg, C.~Bambi, J.~Jiang, H.~Liu and M.~Zhou,
{\it Testing General Relativity with NuSTAR data of Galactic Black Holes},
Astrophys. J. \textbf{913}, 79 (2021),
\href{https://doi.org/10.3847/1538-4357/abf6cd}{https://doi.org/10.3847/1538-4357/abf6cd}
[arXiv:2012.10669 [astro-ph.HE]].

\bibitem{Das:2026zyt}
D.~Das, S.~Shashank and C.~Bambi,
{\it Improved Constraints on Non-Kerr Deviations from Binary Black Hole Inspirals Using GWTC-4 Data},
Class. Quant. Grav. \textbf{43}, 137001 (2026),
\href{https://doi.org/10.1088/1361-6382/ae8118}{https://doi.org/10.1088/1361-6382/ae8118}
[arXiv:2604.15965 [gr-qc]].

\bibitem{Zhao:2023erf}
S.~Zhao, S.~Riaz and C.~Bambi,
{\it About the ability of agnostic x-ray tests of the Kerr hypothesis to discover new physics},
Phys. Rev. D \textbf{109}, 064059 (2024), 
\href{https://doi.org/10.1103/PhysRevD.109.064059}{https://doi.org/10.1103/PhysRevD.109.064059}
[arXiv:2310.18962 [gr-qc]].

\bibitem{Zhu:2019trp}
J.~P.~Zhu, B.~Zhang and Y.~P.~Yang,
{\it Relativistic Astronomy. II. In-Flight Solution of Motion and Test of Special Relativity Light Aberration},
Astrophys. J. \textbf{877}, 14 (2019),
\href{https://doi.org/10.3847/1538-4357/ab1650}{https://doi.org/10.3847/1538-4357/ab1650}
[arXiv:1904.02056 [astro-ph.HE]].

\bibitem{Pretorius:2007jn}
F.~Pretorius and D.~Khurana,
{\it Black hole mergers and unstable circular orbits},
Class. Quant. Grav. \textbf{24}, S83-S108 (2007),
\href{https://doi.org/10.1088/0264-9381/24/12/S07}{https://doi.org/10.1088/0264-9381/24/12/S07}
[arXiv:gr-qc/0702084 [gr-qc]].

\bibitem{Bambi:2014koa}
C.~Bambi, D.~Malafarina and N.~Tsukamoto,
{\it Note on the effect of a massive accretion disk in the measurements of black hole spins},
Phys. Rev. D \textbf{89}, 127302 (2014),
\href{https://doi.org/10.1103/PhysRevD.89.127302}{https://doi.org/10.1103/PhysRevD.89.127302}
[arXiv:1406.2181 [gr-qc]].


\end{thebibliography}
\end{document}